\documentclass{jpp}
\usepackage{graphicx, color}

\usepackage[utf8]{inputenc}
\usepackage[T1]{fontenc}
\usepackage{amsmath,comment}

\newcommand{\vect}[1]{\mbox{\boldmath $#1$}}
\newcommand{\LS}{LS}
\newcommand{\etabar}{\bar{\eta}}

\newcommand{\sgn}{\mathrm{sgn}}
\newcommand{\changed}[1]{{#1}}
\newcommand{\changedd}[1]{{#1}}

\newcommand{\elong}{\kappa_{\mathrm{e}}}

\shorttitle{Figures of merit for stellarators near the magnetic axis}
\shortauthor{M Landreman}

\title{Figures of merit for stellarators near the magnetic axis}

\author{Matt Landreman\aff{1}
  \corresp{\email{mattland@umd.edu}}
 }

\affiliation{\aff{1}Institute for Research in Electronics and Applied Physics,
University of Maryland,
College Park, MD 20742, USA
}

\begin{document}

\maketitle

\begin{abstract}
A new paradigm for rapid stellarator configuration design
has been recently demonstrated, in which the shapes of
quasisymmetric or omnigenous flux surfaces are computed directly using an expansion in small distance
from the magnetic axis. To further develop this approach,
here we derive several other quantities of interest that can be rapidly computed from this near-axis expansion. First, the $\nabla\vect{B}$ and $\nabla\nabla\vect{B}$ tensors are computed, which can be used for direct derivative-based optimization of electromagnetic coil shapes to achieve the desired magnetic configuration. 
Moreover, if the norm of these tensors is large compared to the field strength for a given magnetic field, the field must have a short length scale, suggesting it may be hard to produce with coils that are suitably far away.
Second, we evaluate the minor radius at which the flux surface shapes would become singular, providing a lower bound on the achievable aspect ratio. 
This bound is also shown to be related to an equilibrium beta limit.
Finally, for configurations that are constructed to achieve a desired magnetic field strength to first order in the expansion, we compute the error field that arises due to second order terms.
\end{abstract}


\section{Introduction}

The geometry of a stellarator's magnetic field needs to be
designed to achieve good 
confinement, magnetohydrodynamic stability, and other 
desirable properties.
In recent stellarator experiments such as
HSX and W7-X, this design was performed by optimizing
an objective function computed from numerical magnetohydrodynamic (MHD) equilibrium solutions. 
This approach suffers from the fact that the optimization
algorithm can get trapped in local minima of the objective function, and relatively little insight is gained into the space
of solutions. To address these issues, 
we have recently developed a complementary approach
based on a high aspect ratio approximation \citep{PaperI, PaperII,PaperIII,fitToGarrenBoozer,r2GarrenBoozer, JorgeArbitraryOrder, JorgeQS}.
In this complementary approach, the three-dimensional shape of magnetic surfaces
are directly constructed to achieve a desired magnetic
field strength in Boozer coordinates.
This `near-axis construction' method is
based on an expansion devised by \cite{GB1,GB2}.
The reduced equations 
can be solved many orders of magnitude faster than
the three-dimensional MHD equilibrium solution required at each iteration
of traditional stellarator optimization.

Past work on the near-axis construction has
focused on neoclassical confinement, through the 
properties of quasisymmetry and omnigenity \citep{PaperII,PaperIII,r2GarrenBoozer}.
Recently it has been shown that within this approach, magnetic well and Mercier stability can be computed \citep{LandremanMercier}. One can also evaluate the geometric quantities entering the gyrokinetic equation for microinstabilities and turbulence \citep{Jorge2020}.
In the present paper, we show how a number
of other useful quantities can be computed
directly from a solution of the near-axis equations.
In future work, these figures of merit could be targeted during optimization within the space of near-axis solutions. Such an optimization would have the advantage that the objective function could be evaluated orders of magnitude faster than in traditional stellarator optimization based on full 3D equilibrium calculations. 
The figures of merit derived here could also be applied to high-resolution scans over parameter space, which are made possible by the speed of the near-axis approach. Specifically, during such a scan over near-axis configuration parameters, the figures of merit here could be used to identify regions of parameter space that are uninteresting due to the need for close electromagnetic coils, unreasonably high aspect ratio, or large errors in quasisymmetry. Attention could then be focused on the remaining regions of parameter space.

The first quantities we will evaluate,
in section \ref{sec:tensors}, are the tensors $\nabla\vect{B}$
and $\nabla \nabla\vect{B}$ along the magnetic axis, where $\vect{B}$ is the magnetic field.
These tensors are useful for two reasons. First,
electromagnetic coils can be designed to produce a field matching these tensors,
which (for vacuum fields) means they produce the desired stellarator configuration \citep{Giuliani}. Second, these tensors encode all the scale lengths in
the magnetic field (up to the order of interest), reflecting
how far away the coils may be from the axis. Configurations with short scale lengths in $\vect{B}$ are likely to require very close coils, which is undesirable.
Next, in section \ref{sec:singularity},
we compute the maximum minor radius for which the constructed
surface shapes are smooth and nested.
This critical minor radius is equivalent to a minimum
aspect ratio, beyond which the near-axis construction
does not give physical surface shapes.
Configurations for which this minimum aspect ratio is large
can then be rejected.
We also show in section \ref{sec:betaLimit} that this limit on the aspect ratio is related to an equilibrium limit on $\beta$, the ratio of plasma to magnetic pressure.
Finally, in section \ref{sec:errorField},
we consider the case in which the surface shapes
are constructed to give some desired pattern of
$B$ in Boozer coordinates to first order (in inverse aspect ratio).
The ``error field'' associated with the $B$ at next order is computed.
This information could enable
the parameters of the near-axis equations to be optimized
so that this error field is small.

All numerical calculations shown in this paper were carried out using the code in \cite{zenodoCode}, with data available in \cite{zenodoData}.

\section{Notation}

We employ the `inverse expansion' of \cite{GB1,GB2},
while using identical notation to \cite{r2GarrenBoozer}, hereafter denoted \LS. Several
key definitions are repeated here for convenience.
In Boozer coordinates, the magnetic field has the forms
\begin{align}
\label{eq:BoozerCoords}
\vect{B} = &\nabla\psi \times\nabla\theta + \iota \nabla\varphi \times\nabla\psi, \\
 = &\beta \nabla\psi + I \nabla\theta + G \nabla\varphi, \nonumber
\end{align}
where $\theta$ and $\varphi$ are the poloidal and toroidal Boozer angles,
$2\pi\psi$ is the toroidal flux, 
$I$ and $G$ are constant on $\psi$ surfaces,
and $\iota(\psi)$ is the rotational transform. To facilitate calculations with quasi-helical symmetry, it is convenient to introduce a helical angle
$\vartheta = \theta - N \varphi$, where $N$ is a constant integer that can be set to zero if not considering quasi-helical symmetry. Then
\begin{align}
\vect{B} = &\nabla\psi \times\nabla\vartheta + \iota_N \nabla\varphi \times\nabla\psi,
\label{eq:straight_field_lines_h}
\\
 =& \beta \nabla\psi + I \nabla\vartheta + (G+NI) \nabla\varphi,
\label{eq:Boozer_h}
\end{align}
where $\iota_N = \iota - N$.

The position vector $\vect{r}$ at an arbitrary point can be written
\begin{align}
\label{eq:positionVector}
\vect{r}(r,\vartheta,\varphi) = \vect{r}_0(\varphi)
+X(r,\vartheta,\varphi) \vect{n}(\varphi)
+Y(r,\vartheta,\varphi) \vect{b}(\varphi)
+Z(r,\vartheta,\varphi) \vect{t}(\varphi),
\end{align}
where $\vect{r}_0(\varphi)$ is the position vector along the magnetic axis. The Frenet-Serret unit vectors of the axis $(\vect{t},\vect{n},\vect{b})$ satisfy
\begin{align}
\frac{d\varphi}{d\ell}
\frac{d\vect{r}_0}{d\varphi} = \vect{t}, 
\hspace{0.3in}
\frac{d\varphi}{d\ell}
\frac{d\vect{t}}{d\varphi} = \kappa \vect{n}, 
\hspace{0.3in}
\frac{d\varphi}{d\ell}
\frac{d\vect{n}}{d\varphi} = -\kappa \vect{t} + \tau \vect{b}, 
\hspace{0.3in}
\frac{d\varphi}{d\ell}
\frac{d\vect{b}}{d\varphi} = -\tau \vect{n}, 
\label{eq:Frenet}
\end{align}
and $\vect{t}\times\vect{n}=\vect{b}$. Here, $\kappa(\varphi)$ is the axis curvature, and $\tau(\varphi)$ is the axis torsion, 
and $\ell$ is the arclength along the axis. 

The expansion about the magnetic axis is developed by writing
\begin{align}
X(r,\vartheta,\varphi)
= r X_1(\vartheta,\varphi) + r^2 X_2(\vartheta,\varphi) + r^3 X_3(\vartheta,\varphi) + \ldots,
\label{eq:radial_expansion}
\end{align}
where the effective minor radius coordinate $r$ is defined via $2 \pi \psi = \pi r^2 \bar{B}$, and where the constant $\bar{B}$ is a reference field strength. 
The quantities $Y$ and $Z$ in (\ref{eq:positionVector}) 
are expanded analogously. 
All scale lengths in the system except $r$ are ordered as comparable to a large length scale $\mathcal{R}$,
from which we define the expansion parameter $\epsilon = r / \mathcal{R}$.
Then (\ref{eq:radial_expansion})
represents an expansion in $\epsilon \ll 1$.
We expand $B$ and $\beta$ in a similar way to (\ref{eq:radial_expansion}) but with an $O(\epsilon^0)$ term:
\begin{align}
\label{eq:radial_expansion_B}
B(r,\vartheta,\varphi)
= B_0(\varphi) + r B_1(\vartheta,\varphi) + r^2 B_2(\vartheta,\varphi)+ r^3 B_3(\vartheta,\varphi) + \ldots.
\end{align}
The radial profiles
$G(r)$, $I(r)$, $p(r)$, and $\iota_N(r)$ have expansions with only even powers of $r$, since they must be even in $r$:
\begin{align}
\label{eq:radial_expansion_G}
G(r) = G_0 + r^2 G_2 + r^4 G_4 + \ldots.
\end{align}
In the analogous expansion of $I(r)$, $I_0 = 0$ since $I(r)$ is proportional to the toroidal current inside the surface $r$.
Considering that physical quantities should be analytic at
the magnetic axis, the expansion coefficients have the form
\begin{align}
\label{eq:poloidal_expansions}
X_1(\vartheta,\varphi) = &X_{1s}(\varphi) \sin(\vartheta) + X_{1c}(\varphi) \cos(\vartheta), \\
X_2(\vartheta,\varphi) = &X_{20}(\varphi) + X_{2s}(\varphi) \sin(2\vartheta) + X_{2c}(\varphi) \cos(2\vartheta), \nonumber \\
X_3(\vartheta,\varphi) = &X_{3s3}(\varphi) \sin(3\vartheta) + X_{3s1}(\varphi) \sin(\vartheta) + X_{3c3}(\varphi) \cos(3\vartheta) + X_{3c1}(\varphi) \cos(\vartheta).
 \nonumber
\end{align}
(A detailed discussion is given in appendix A of \cite{PaperI}.)
The quantities $Y$, $Z$, $B$, and $\beta$ have expansion coefficients of the same form as (\ref{eq:poloidal_expansions}).

\changed{Next, we apply the dual relations, $\nabla r=(\partial\vect{r}/\partial\vartheta\times\partial\vect{r}/\partial\varphi)/\sqrt{g}$ and cyclic permutations, where $\sqrt{g}=(\nabla r\cdot\nabla\vartheta\times\nabla\varphi)^{-1}=(G+\iota I)r\bar{B}/B^2$ is the Jacobian of the $(r,\vartheta,\varphi)$ coordinates.}
Derivatives of the position vector (\ref{eq:positionVector}) are substituted into the dual relations 
to obtain the vectors $\nabla r$, $\nabla\vartheta$, and $\nabla\varphi$. These gradient vectors are then substituted into (\ref{eq:straight_field_lines_h}) $=$ (\ref{eq:Boozer_h}) and $(\nabla\times\vect{B})\times\vect{B} = \mu_0 \nabla p$. The  condition $B(r,\vartheta,\varphi)=B(r,\vartheta)$ can also be imposed if one desires quasisymmetry. Equations are thereby obtained at each order in $\epsilon \ll 1$. These equations can be found in the appendix of \cite{GB2} and appendix A of \LS.

We note two signs that appear in the analysis: $s_G = \sgn(G) = \pm 1$, and $s_\psi = \sgn(\psi) = \sgn(\bar{B}) = \pm 1$. 
\changed{As discussed in appendix A.1 of \cite{LandremanMercier}, these signs are associated with the choice of $\varphi$ and $\theta$: each sign can be flipped by reversing the direction in which these angles increase.}
If one is considering quasisymmetry, the normalizing field can be taken to be $\bar{B} = s_\psi B_0$.

With the notation and expansion now defined, we proceed to
derive the new figures of merit that can be computed
from these expansion coefficients.


\section{$\nabla\vect{B}$ and $\nabla\nabla\vect{B}$ tensors}
\label{sec:tensors}



There are several reasons why it is valuable to evaluate the tensors $\nabla\vect{B}$ and $\nabla\nabla \vect{B}$ on the magnetic axis in terms
of the near-axis expansion coefficients. First, 
consistency between these tensors and the magnetic field from the Biot-Savart formula can be achieved by numerical optimization, enabling direct optimization of coil shapes for quasisymmetry. 
Demonstrations of this method for $O(\epsilon)$ quasisymmetry 
are presented in \cite{Giuliani},
in which the formula for $\nabla\vect{B}$ derived here is used.
It may be possible to optimize coils for $O(\epsilon^2)$ quasisymmetry (requiring $\nabla\nabla \vect{B}$) in future work.

Second, $\nabla\vect{B}$ and $\nabla\nabla \vect{B}$ provide information about how feasible it is to generate a given near-axis stellarator solution using \emph{distant} magnets.
This is because $\nabla\vect{B}$
encodes all the (inverse) scale lengths in the magnetic field that can be known from a $O(\epsilon^1)$ solution, and $\nabla\nabla \vect{B}$ encodes all the additional (inverse) scale lengths in the magnetic field that can be known from a $O(\epsilon^2)$ solution. It is unlikely that magnets can be much farther from the axis than any of these scale lengths.
\changed{This is because in magnetostatics the field from a small-scale current structure decays rapidly as one moves away from the current; for example in slab geometry a steady sheet current $\propto \sin kx$ with wavenumber $k$ on the $z=0$ plane produces a vacuum field $\propto \exp(-|kz|)$ that is exponentially small beyond distances $|z| > 1/|k|$.}
Note that these scale lengths associated with $\nabla\vect{B}$ and $\nabla\nabla\vect{B}$ are not directly related to  scale lengths in magnetic surface shapes: surface shapes near an X-point have a radius of curvature that shrinks to zero, but the scale lengths in $\vect{B}$ remain nonzero on a separatrix. We therefore expect scale lengths derived from $\nabla\vect{B}$ and $\nabla\nabla\vect{B}$ to be better indicators of the distance to a coil than scale lengths in the surface shape.
`Worst-case' scale lengths can be constructed by summing the squares of the matrix elements in these tensors, as in
\begin{align}
\label{eq:distance_to_wire}
L_{\nabla B} &= B \sqrt{2 / ||\nabla\vect{B}||^2},
\\
\label{eq:grad_grad_B_scale_length}
L_{\nabla\nabla B} &= \sqrt{4B / ||\nabla\nabla\vect{B}||
},
\end{align}
\changed{for the (squared) Frobenius norm} $||\nabla \vect{B}||^2=\sum_{i,j=1}^3 (\nabla\vect{B})_{i,j}^2$ and $||\nabla\nabla\vect{B}||^2 = \sum_{i,j,k=1}^3 (\nabla\nabla\vect{B})_{i,j,k}^2$.
These expressions are motivated at the end of this section
by considering an infinite straight wire, with $L_{\nabla B}$ and $L_{\nabla\nabla B}$ both constructed to give the distance from the wire.
Since it is important in stellarators to maximize the plasma-to-coil distance, we expect that configurations with small $L_{\nabla B}$ or $L_{\nabla\nabla B}$ can be excluded as impractical.
Both this application and the application in \cite{Giuliani} motivate a calculation of $\nabla \vect{B}$ and $\nabla\nabla\vect{B}$ in terms of the Garren-Boozer expansion.

To proceed, we first evaluate the vector $\vect{B}$ 
near the axis. This can be done using (\ref{eq:straight_field_lines_h}) in the form
\begin{align}
\label{eq:vectorB}
\vect{B} = \frac{B^2}{G+\iota I} \left[ 
\left(\Lambda + \iota_N \frac{\partial Z}{\partial\vartheta}\right)\vect{t}
+\left(\Xi + \iota_N \frac{\partial X}{\partial\vartheta}\right)\vect{n}
+\left(\Upsilon + \iota_N \frac{\partial Y}{\partial\vartheta}\right)\vect{b} \right],
\end{align}
where
\begin{align}
\Lambda = \ell' + \frac{\partial Z}{\partial\varphi} - X \ell' \kappa 
, \hspace{0.3in}
\Xi = \frac{\partial X}{\partial\varphi} - Y \ell' \tau + Z \ell' \kappa
, \hspace{0.3in}
\Upsilon = \frac{\partial Y}{\partial\varphi} + X \ell' \tau
\end{align}
following notation from \cite{GB1},
with $\ell'=|\partial\vect{r}_0/\partial\varphi|$.
Notice $X \sim Y \sim \epsilon^1$, $Z \sim \epsilon^2$,
$\Lambda \sim \epsilon^0$, and $\Xi \sim \Upsilon \sim \epsilon^1$.
Writing $\vect{B} = \vect{B}_0 + r \vect{B}_1 + r^2 \vect{B}_2 + \ldots$, and noting $\ell' = |G_0| / B_0$, the leading term in (\ref{eq:vectorB}) is 
\begin{align}
\label{eq:B0}
\vect{B}_0 = s_G B_0 \vect{t}.
\end{align}
The terms of next order in (\ref{eq:vectorB}) give
\begin{align}
\label{eq:B1}
\vect{B}_1 = \frac{B_0^2}{G_0} \left[
X_1 \ell' \kappa \vect{t}
+\left( \frac{\partial X_1}{\partial\varphi}-Y_1 \ell' \tau + \iota_{N0} \frac{\partial X_1}{\partial\vartheta}\right)\vect{n}
+\left(\frac{\partial Y_1}{\partial\varphi} + X_1 \ell' \tau + \iota_{N0} \frac{\partial Y_1}{\partial\vartheta}\right)\vect{b} \right],
\end{align}
where $B_1 = \kappa B_0 X_1$ has been employed.

We next use the dual relations to write
\begin{align}
\label{eq:gradB_setup}
\nabla \vect{B} &=  (\nabla r)\frac{\partial \vect{B}}{\partial r}
+ (\nabla\vartheta)\frac{\partial \vect{B}}{\partial\vartheta}
+ (\nabla\varphi)\frac{\partial \vect{B}}{\partial\varphi}
\\
&= \frac{B^2}{(G+\iota I)r\bar{B}} \left[
 \left( \frac{\partial X}{\partial\vartheta}\vect{n}+\frac{\partial Y}{\partial\vartheta}\vect{b}+\frac{\partial Z}{\partial\vartheta}\vect{t}\right)
\times \left( \Lambda\vect{t}+\Xi\vect{n}+\Upsilon\vect{b}\right) 
\frac{\partial \vect{B}}{\partial r}\right. \nonumber \\
& \hspace{0.1in}
+ 
\left( \Lambda\vect{t}+\Xi\vect{n}+\Upsilon\vect{b}\right) \times
\left( \frac{\partial X}{\partial r}\vect{n}+\frac{\partial Y}{\partial r}\vect{b}+\frac{\partial Z}{\partial r}\vect{t}\right) 
\frac{\partial \vect{B}}{\partial\vartheta}
\nonumber \\
& \hspace{0.1in}\left. 
\changed{+}\left( \frac{\partial X}{\partial r}\vect{n}+\frac{\partial Y}{\partial r}\vect{b}+\frac{\partial Z}{\partial r}\vect{t}\right) \times
\left( \frac{\partial X}{\partial\vartheta}\vect{n}+\frac{\partial Y}{\partial\vartheta}\vect{b}+\frac{\partial Z}{\partial\vartheta}\vect{t}\right)
\frac{\partial \vect{B}}{\partial\varphi}
\right].
\nonumber
\end{align}
The terms of $O(\epsilon^0)$ are
\begin{align}
\nabla \vect{B} \approx &
\frac{B_0^2}{G_0 \bar{B}} \left[
\ell' \left(-\frac{\partial X_1}{\partial\vartheta}\vect{b} + \frac{\partial Y_1}{\partial\vartheta}\vect{n}\right) \vect{B}_1
+ \ell' (X_1 \vect{b} - Y_1 \vect{n}) 
\frac{\partial \vect{B}_1}{\partial\vartheta}
\right. \nonumber \\
& \hspace{0.5in}
\left.+ \left( X_1 \frac{\partial Y_1}{\partial \vartheta} - Y_1 \frac{\partial X_1}{\partial \vartheta}\right) \vect{t} 
\frac{d \vect{B}_0}{d \varphi}\right],
\label{eq:gradB_intermediate}
\end{align}
where in the last term, (\ref{eq:B0}) gives $d \vect{B}_0/d\varphi = s_G B'_0 \vect{t} + s_G B_0 \ell' \kappa \vect{n}$. 
We substitute in (\ref{eq:B1}),
 noting that for any functions
\begin{align}
P(\vartheta) = P_s \sin\vartheta + P_c \cos\vartheta,
\hspace{0.5in}
Q(\vartheta) = Q_s \sin\vartheta + Q_c \cos\vartheta,
\end{align}
we have
\begin{align}
P \frac{\partial Q}{\partial\vartheta} - Q \frac{\partial P}{\partial \vartheta} = P_c Q_s - P_s Q_c
,\hspace{0.5in}
PQ + \frac{\partial P}{\partial\vartheta}\frac{\partial Q}{\partial\vartheta} = P_s Q_s + P_c Q_c,
\end{align}
and $\partial^2 P/\partial\vartheta^2 = -P$.
We also note (A21) in \LS, $X_{1c}Y_{1s}-X_{1s}Y_{1c}=s_G \bar{B}/B_0$
\changed{(the condition that the first-order flux surface enclose the proper toroidal flux at each $\varphi$.)}
We thereby obtain the final result for a general configuration:
\begin{align}
\label{eq:gradB_general}
\nabla \vect{B} \approx &\frac{B_0^2}{\bar{B} \ell'} 
\left[ \left( X'_{1c} Y_{1s} - X'_{1s} Y_{1c} + \iota_{N0} [X_{1s} Y_{1s} + X_{1c} Y_{1c}]\right) \vect{n}\vect{n}
\vphantom{\frac{\bar{B}}{B_0}}
\right. 
\\
&  \hspace{0.0in}+  \left(X_{1c} X'_{1s} - X_{1s} X'_{1c} - \frac{s_G \bar{B} \ell' \tau}{B_0} - \iota_{N0} [X_{1s}^2+X_{1c}^2]\right) \vect{b}\vect{n}
 \nonumber \\
& \hspace{0.0in}+\left( Y'_{1c}Y_{1s} - Y'_{1s}Y_{1c} + \frac{s_G \bar{B} \ell' \tau}{B_0} + \iota_{N0}[Y_{1s}^2 + Y_{1c}^2]\right) \vect{n}\vect{b} 
\nonumber \\
& \hspace{0.0in} \left. \vphantom{\frac{\bar{B}}{B_0}}
+\left(X_{1c} Y'_{1s} - X_{1s} Y'_{1c} - \iota_{N0} [X_{1s}Y_{1s}+X_{1c}Y_{1c}]\right) \vect{b}\vect{b} \right] 
+ s_G B_0 \kappa (\vect{t}\vect{n} + \vect{n}\vect{t})
+ \frac{ s_G B'_0}{\ell'} \vect{t}\vect{t}.
\nonumber
\end{align}
Using (A23) in \LS, \changed{(which expresses how on-axis rotational transform is driven by rotating elongation, axis torsion, and toroidal current),} it can be seen that the $\vect{n}\vect{b}$ component
and the $\vect{b}\vect{n}$ component become equal when $I_2=0$,
giving the expected symmetry of $\nabla\vect{B}$ for a vacuum field.

In the case of quasisymmetry, (\ref{eq:gradB_general}) simplifies due to $B'_0=0$ and $X_{1s}=0$, and we also have $\bar{B}=s_\psi B_0$. Thus, the $\nabla\vect{B}$ tensor for quasisymmetry is
\begin{align}
\label{eq:gradB_QS}
\nabla \vect{B} \approx &\frac{s_\psi B_0}{ \ell'} 
\left[ \left( X'_{1c} Y_{1s}  + \iota_{N0} X_{1c} Y_{1c}\right) \vect{n}\vect{n}
+\left( - s_G s_\psi  \ell' \tau - \iota_{N0} X_{1c}^2\right) \vect{b}\vect{n}
\right.
\\
& \hspace{0.5in}+\left( Y'_{1c}Y_{1s} - Y'_{1s}Y_{1c} + s_G s_\psi  \ell' \tau + \iota_{N0}[Y_{1s}^2 + Y_{1c}^2]\right) \vect{n}\vect{b} 
\nonumber \\
&  \hspace{0.5in} \left. +  
\left(X_{1c} Y'_{1s}  - \iota_{N0} X_{1c}Y_{1c}\right) \vect{b}\vect{b} \right] 
+ s_G B_0 \kappa (\vect{t}\vect{n} + \vect{n}\vect{t}).
\nonumber
\end{align}

Analogously to (\ref{eq:gradB_setup})-(\ref{eq:gradB_intermediate}), we can evaluate the $\nabla \nabla \vect{B}$ tensor as well.
To leading order,
\begin{align}
\nabla \nabla  \vect{B} \approx &
\frac{B_0^2}{G_0 \bar{B}} \left[
\ell' \left(-\frac{\partial X_1}{\partial\vartheta}\vect{b} + \frac{\partial Y_1}{\partial\vartheta}\vect{n}\right) (\nabla \vect{B})_1
+ \ell' (X_1 \vect{b} - Y_1 \vect{n}) 
\frac{\partial (\nabla \vect{B})_1}{\partial\vartheta}
\right. \nonumber \\
& \hspace{0.5in}
\left.+ \left( X_1 \frac{\partial Y_1}{\partial \vartheta} - Y_1 \frac{\partial X_1}{\partial \vartheta}\right) \vect{t} 
\frac{\partial (\nabla \vect{B})_0}{\partial \varphi}\right].\label{eq:nablanablab1}
\end{align}
The result is too lengthy to write here, but it can be found in a Mathematica notebook in \cite{zenodoCode}.
The result depends on $X_2$, $Y_2$, and $Z_2$.

For reference, we note the values of $\nabla\vect{B}$ and
$\nabla\nabla\vect{B}$ in the limit of an axisymmetric \changed{and purely toroidal} vacuum field, which is the field of an infinite straight 
\changed{vertical} wire  \changed{at major radius $R=0$.} This limit can be used for verification and to derive figures of merit such as (\ref{eq:distance_to_wire})-(\ref{eq:grad_grad_B_scale_length}), the equivalent distance to a coil. The field is $\vect{B} =  s_G \mu_0 I \vect{t} / (2 \pi R)$, where $I>0$ is the current in the wire, $R$ is the distance from the wire,
\changed{
and $\vect{t}=\vect{e}_\phi$ is the unit vector in the direction of the standard cylindrical angle $\phi$. Note also $\vect{n}=-\vect{e}_R$.} Then
\begin{align}
\nabla\vect{B} = (\nabla R)\frac{\partial\vect{B}}{\partial R} + (\nabla\phi)\frac{\partial\vect{B}}{\partial\phi}
=\frac{s_G \mu_0 I}{2 \pi R^2}(\vect{n}\vect{t} + \vect{t}\vect{n}),
\end{align}
consistent with the appropriate limit of (\ref{eq:gradB_QS}) ($d/d\varphi=0$, $\tau=0$, $\iota_{N0}=0$.) Forming the squared Frobenius norm of this expression and solving for $R$ gives 
$R = L_{\nabla B}$ where $L_{\nabla B}$ is
the figure of merit (\ref{eq:distance_to_wire}). Similarly,
\begin{align}
\nabla\nabla\vect{B} = (\nabla R)\frac{\partial(\nabla\vect{B})}{\partial R} + (\nabla\phi)\frac{\partial(\nabla\vect{B})}{\partial\phi}
=\frac{s_G \mu_0 I}{\pi R^3}(\vect{t}\vect{n}\vect{n} + \vect{n}\vect{t}\vect{n} + \vect{n}\vect{n}\vect{t} - \vect{t}\vect{t}\vect{t}).
\end{align}
Squaring and solving for $R$ gives $R = L_{\nabla\nabla B}$ where $L_{\nabla \nabla B}$ is the expression in (\ref{eq:grad_grad_B_scale_length}).

\changed{
For this case of a purely toroidal $B\propto 1/R$ field, the same $\vect{B}$ can be produced in an axisymmetric toroidal domain by a poloidal sheet current on the domain boundary. Hence the currents can be moved closer to an evaluation point at some fixed $R>0$ without changing $L_{\nabla B}$ or $L_{\nabla\nabla B}$. This example illustrates that for a given field, there is no \emph{unique} distance to a coil. The best we can hope to compute from a given $\vect{B}$ is a \emph{maximum} distance to a current, reflecting the distance to a singularity in the exterior \citep{GreeneSingularity}. In this axisymmetic example, where there must be current through the donut hole of the toroidal region where $B\propto 1/R$, $L_{\nabla B}$ and $L_{\nabla\nabla B}$ give the correct maximum possible distance to a current.
}

\begin{figure}
  \centering
  \includegraphics[width=4.8in]{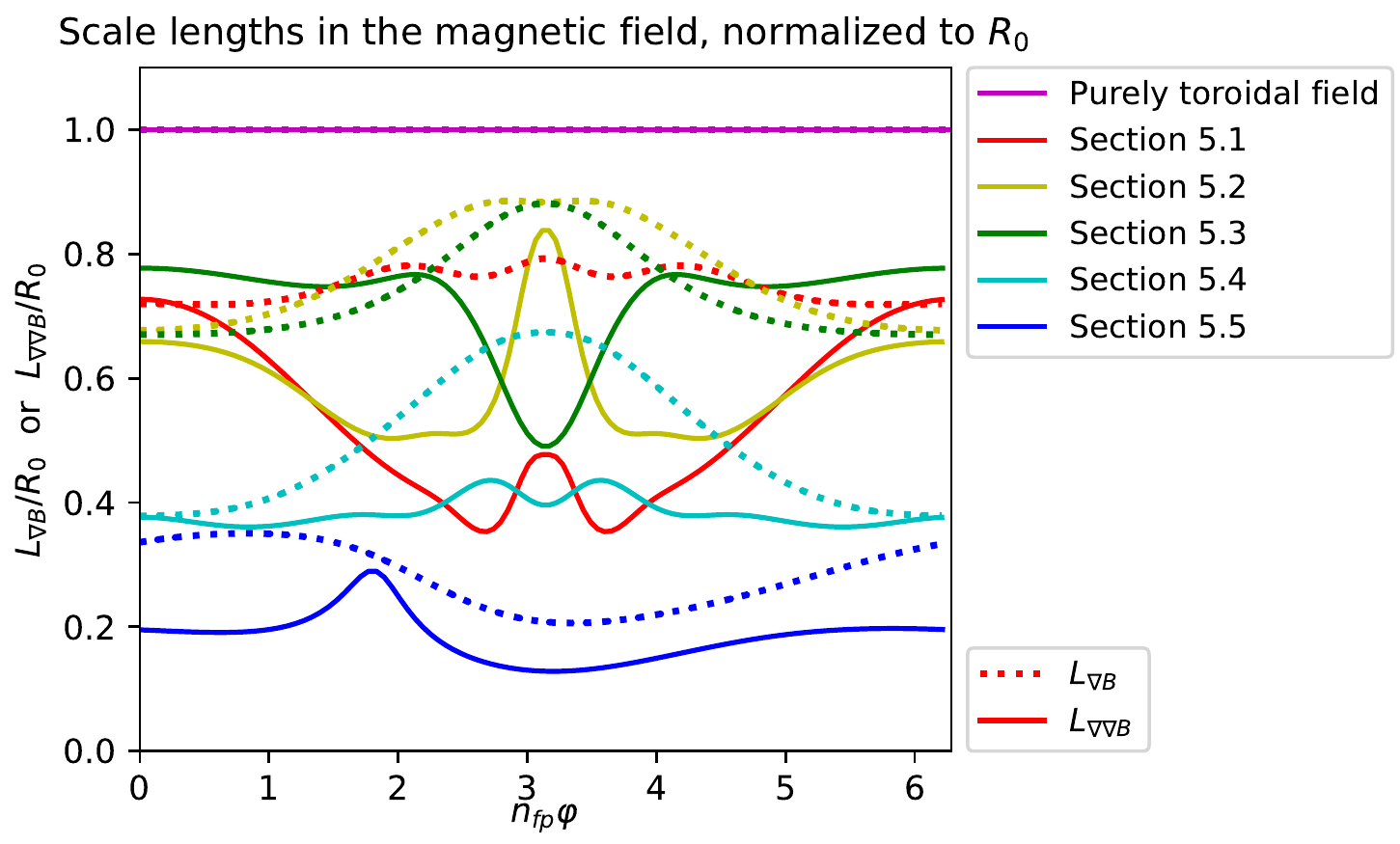}
  \caption{
Scale lengths in the magnetic field, indicating
measures of the effective distance to an electromagnetic coil,
evaluated for the five examples in \cite{r2GarrenBoozer}. 
\changed{Section numbers in the legend refer to this earlier paper.}
Both of the scale lengths (\ref{eq:distance_to_wire}) and (\ref{eq:grad_grad_B_scale_length}) are shown.
}
\label{fig:gradBTensor}
\end{figure}

Figure \ref{fig:gradBTensor} shows the scale lengths
$L_{\nabla B}$ and $L_{\nabla\nabla B}$ for the five $O(\epsilon^2)$ quasisymmetric configurations
discussed in \LS.
\changed{Colors and legend text in the figure refer to section numbers in \LS, each representing a different stellarator configuration.}
The scale lengths are normalized to the major radius to yield dimensionless figures of merit. By design, this normalized measure is 1 for a purely toroidal field, and small values
are undesirable. The examples from sections 5.1-5.3 are quasi-axisymmetric, while the examples from sections 5.4-5.5 are quasi-helically symmetric. The example of section 5.5 lacks
stellarator symmetry, while the other examples are stellarator-symmetric. 
\changed{Therefore the two curves for this example lack reflection symmetry about $n_{fp}\varphi=\pi$, whereas the curves for the stellarator-symmetric examples do have this reflection symmetry.}
The example of section 5.5 has quite strong shaping, as is evident in figure 17 of \LS. Therefore it is not surprising that this example has significantly smaller values for the normalized scale lengths compared to the other examples.
This initial evidence is at least suggestive that $L_{\nabla B}$
and $L_{\nabla\nabla B}$ may be useful as proxies for the complexity of a magnetic configuration. 
\changedd{Specifically, noting that $L_{\nabla B}$ and $L_{\nabla \nabla B}$ are somewhat independent, and that the shortest scale length dominates, we conjecture that small values of $\min(L_{\nabla B}, \, L_{\nabla \nabla B})$ are correlated with high coil complexity.}
Further study involving coil optimization is needed to verify this conjecture. 
Note that 
$L_{\nabla B}$
and $L_{\nabla\nabla B}$ can be evaluated in under 1 ms for reasonable resolutions, roughly 4 orders of magnitude faster than even the fastest calculations of coil shapes with current-potential methods like REGCOIL \citep{regcoil}.


\section{Minimum aspect ratio without intersecting surfaces}
\label{sec:singularity}

\begin{figure}
  \centering
  \includegraphics[width=4.8in]{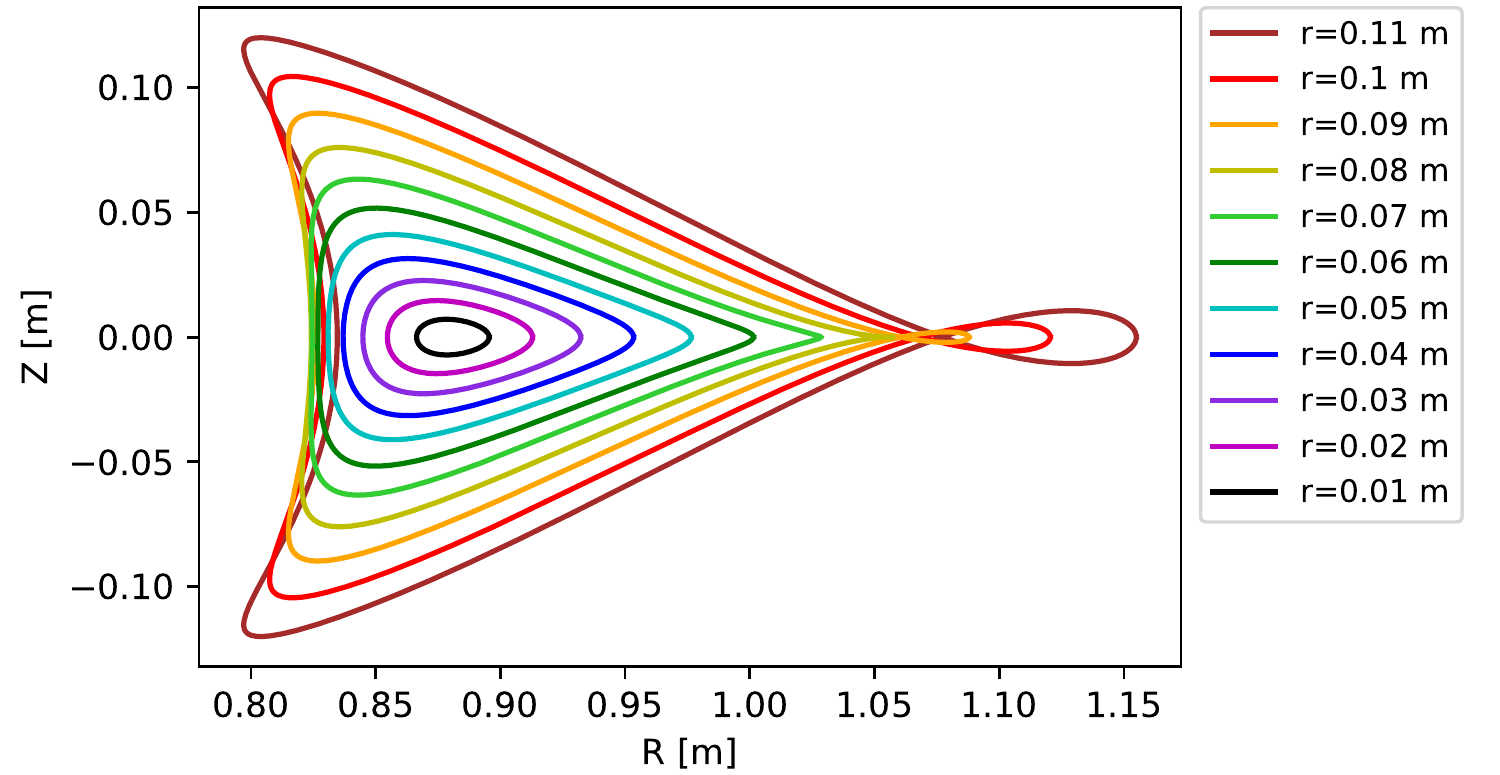}
  \caption{
Several unphysical behaviors can arise in the constructed flux surface shapes
for sufficiently large minor radius $r$.
In the example here, showing the $\phi=0$ cross-section
of the configuration of section \ref{sec:singularityExample},
each surface intersects itself on the large-$R$ side when $r>0.085$ m, and the surfaces are not properly nested on the small-$R$ side for $r>0.077$ m.
}
\label{fig:singularity}
\end{figure}

When a stellarator configuration is constructed to $O(\epsilon^2)$
using the near-axis expansion, a common problem is the following.
The triangularity of the constructed boundary surface grows
with $r$, as does the shift of the surface centroid compared to the magnetic axis. For $r$ beyond a critical value $r_c$, the surface can begin to self-intersect, or the surfaces may no longer be nested,
as shown in figure \ref{fig:singularity}.
These intersections and overlaps of the surfaces are not physically realizable, but rather are an indication
that the small-$r$ expansion has broken down.
These problems with the surface geometry 
put a lower limit on the aspect
ratio at which boundary surfaces can be constructed
by the near-axis method. Therefore, it is useful to be able 
to calculate $r_c$.

In effect, $r_c$ is a convenient summary of how rapidly the
coefficients of the $r$ expansion increase with the order in the expansion. 
\changed{The $(X_j, Y_j, Z_j)$ coefficients tend to increase with $j$, since in the Garren-Boozer equations the order-$j$ terms depend on the $d/d\varphi$ derivatives of the order-$(j-1)$ terms. This increase with $j$ limits the radius of convergence of the expansion in $r$.}
When $r_c$ is small, the coefficients increase rapidly with order so the expansion is accurate only for small $r$. We wish to find solutions of the near-axis equations for which the coefficients do not increase rapidly with order, so the expansion is accurate for relatively large $r$. Hence, we seek solutions with large $r_c$.

For quasisymmetric configurations, this critical value $r_c$
for singularity is also useful as a measure of the accuracy of quasisymmetry.
This is because the construction achieves quasisymmetry through $O(\epsilon^2)$ but not at $O(\epsilon^3)$, so quasisymmetry
is not accurately obtained at values of $r$ for which the $O(\epsilon^3)$ terms matter. 
\changed{One can expect the quantities $\{B_j, X_j, Y_j, Z_j\}$ at each order $j$ to all be roughly comparable in magnitude since they are related by the equations at each order of the expansion, e.g. (A32)-(A48) of \LS.}
The $O(\epsilon^3)$ terms in the
shape evidently matter when $r \sim r_c$, for then the 
$O(\epsilon^2)$ shape is unphysical.
Therefore, configurations with large $r_c$ can be expected to have
good quasisymmetry throughout a larger volume than configurations with small $r_c$.
Motivated by these  reasons above, in this section we seek a method to calculate $r_c$.

Considering the type of singularity on the large-$R$ side of figure
\ref{fig:singularity}, the sharp edge in the surface shape is reminiscent of the X-point in a diverted tokamak. 
It may therefore be possible to take advantage of this kind of singularity to design a stellarator divertor. We will not attempt to pursue this possibility here.

\subsection{Problem formulation}
\label{sec:singularityProblem}

The condition for self-intersection or overlap of the flux surfaces, i.e. constant-$r$ surfaces,
is $\sqrt{g}=0$, where $\sqrt{g}$ is the Jacobian
\begin{equation}
    \sqrt{g}= \frac{\partial \vect{r}}{\partial r} \times 
    \frac{\partial \vect{r}}{\partial\vartheta} \cdot 
    \frac{\partial \vect{r}}{\partial\varphi}.
    \label{eq:sqrtg}
\end{equation}
We define $r_c$ as the minimum positive value of $r$ such that
$\sqrt{g}=0$. Since this definition has the form of a constrained optimization problem, we introduce a Lagrange multiplier $\lambda$,
and seek stationary points of the Lagrangian $\mathcal{L}\changed{(\lambda,r,\vartheta,\varphi)} = r + \lambda \sqrt{g}$. Variation with respect to $\lambda$ recovers the constraint $\sqrt{g}=0$. Variation of $\mathcal{L}$ with respect to the spatial coordinates yields
\begin{align}
    \lambda\,  \partial \sqrt{g} / \partial r = -1,
    \label{eq:d_sqrtg_d_r} \\
    \partial \sqrt{g} / \partial \vartheta = 0,
    \label{eq:d_sqrtg_d_theta} \\
    \partial \sqrt{g} / \partial \varphi = 0.
    \label{eq:d_sqrtg_d_phi}
\end{align}
Eq (\ref{eq:d_sqrtg_d_r}) effectively determines $\lambda$, and will not be needed. In practice, since the shape coefficients $\{X,Y,Z\}$
are available on a grid in $\varphi$, it is convenient to replace (\ref{eq:d_sqrtg_d_phi}) with minimization over the same $\varphi$ grid, as follows. We define $\hat{r}_c(\varphi)$ as the solution of
$\sqrt{g}=0$ and (\ref{eq:d_sqrtg_d_theta}) at given $\varphi$,
we can evaluate $\hat{r}_c(\varphi)$ on the numerical grid in $\varphi$, and we then identify $r_c = \min_\varphi \hat{r}_c(\varphi)$. The key task then is to compute $\hat{r}_c(\varphi)$.

To this end, we substitute the position vector (\ref{eq:positionVector}) and
expansions
(\ref{eq:radial_expansion}) and (\ref{eq:poloidal_expansions})
into (\ref{eq:sqrtg}),  using (\ref{eq:Frenet}).
If the expansion for $\{X,Y,Z\}$ is truncated after $\{X_1,Y_1,Z_1\}$, corresponding to the $O(\epsilon)$ construction, 
then $\sqrt{g} = (1-r X_1 \kappa)r s_G \bar{B} \ell' / B_0$, and
the solution of $\sqrt{g}=0$ and $\partial \sqrt{g}/\partial\vartheta=0$ gives $\hat{r}_c(\varphi) = 1/(\kappa \sqrt{X_{1s}^2 + X_{1c}^2})$.
For the rest of section \ref{sec:singularity} we will  consider the more complicated problem of the $O(\epsilon^2)$ construction,
in which 
the expansion for $\{X,Y,Z\}$ is truncated after $\{X_2,Y_2,Z_2\}$.
In this case,
\changed{the product in (\ref{eq:sqrtg}) involving three copies of the position vector includes}
terms scaling as $r^1$ through $r^5$:
\begin{equation}
\sqrt{g}=r \sum_{j=0}^4 r^j g_j(\vartheta,\varphi)
\end{equation}
where $g_0=g_0(\varphi)$ is independent of $\vartheta$,
\begin{align}
g_1 &= g_{1s}(\varphi) \sin\vartheta + g_{1c}(\varphi) \cos\vartheta, \\
g_2 &= g_{20}(\varphi) + g_{2s}(\varphi) \sin 2\vartheta + g_{2c}(\varphi) \cos 2\vartheta, \; \mathrm{etc.},
\end{align}
analogous to (\ref{eq:radial_expansion}).
We find 
$g_0=(X_{1c}Y_{1s} -X_{1s}Y_{1c})\ell'= s_G \bar{B} \ell' / B_0$
\changed{and}
$g_{1s,c} = -2 s_G \bar{B}\ell' B_{1s,c}/B_0^2$.
\changed{These expressions can be derived using (A21) and (A32)-(A33) in \LS,
which express equality of the $\vect{t}$ components of the covariant and contravariant representations of $\vect{B}$.}
These results for $g_0$ and $g_1$ can also be derived by 
expanding $(G+\iota I)r \bar{B}/ B^2$,
which is equal to $\sqrt{g}$ if all orders
in the expansion are retained.
However, since the expansion is truncated after $\{X_2,Y_2,Z_2\}$,
$g_j$ does not equal the corresponding term in the expansion of $(G+\iota I)r \bar{B}/ B^2$ for $j \ge 2$.
Explicit expressions for $g_2$ are given in Appendix \ref{apx:JacobianCoeffs}.

At each $\varphi$, the equations $\sqrt{g}=0$ and $\partial \sqrt{g}/\partial\vartheta$ then give two equations
for the two unknowns $\vartheta$ and $r=\hat{r}_c$.
In principle this nonlinear system could be solved numerically
using Newton's method.
As with any such nonlinear problem, it is hard to ensure that all possible solutions are found, since the numerical solution depends on an initial guess.
However, we are primarily interested in the smallest positive solution for $\hat{r}_c$, which is the root that sets the limit on minimum aspect ratio. In the next subsection we
show how this smallest positive root can be found more robustly using 
an asymptotic approach. The approximate root location computed in the next subsection can serve as a good initial guess for Newton iteration if a more accurate
value of $r_c$ is desired.


\subsection{Robust solution}
\label{sec:robustSingularity}

Since we seek solutions of $\sqrt{g}=0$ and $\partial\sqrt{g}/\partial\vartheta=0$ for small $r$,
we keep only the leading three orders in $\sqrt{g}$: 
$\sqrt{g} \approx r(g_0+r g_1 + r^2 g_2)$. Then (\ref{eq:d_sqrtg_d_theta})
can be written
\begin{equation}
\label{eq:r}
    r = \frac{g_{1c} \sin\vartheta -g_{1s}\cos\vartheta}{2(g_{2s} \cos 2\vartheta - g_{2c} \sin 2\vartheta)}.
\end{equation}
Using this result to eliminate $r$ in $\sqrt{g}=0$, we obtain
\begin{equation}
\label{eq:K}
    K_0 + K_{2s} \sin 2\vartheta + K_{2c} \cos 2\vartheta
    +K_{4s} \sin 4\vartheta + K_{4c} \cos 4\vartheta=0,
\end{equation}
where
\begin{align}
\label{eq:K0}
K_0 &=
2 g_{20}(g_{1c}^2+g_{1s}^2)+8 g_0(g_{2c}^2+g_{2s}^2)+
3 g_{2c}(g_{1s}^2-g_{1c}^2)
 - 6 g_{1c} g_{1s} g_{2s}, \\
 K_{2s}&=
 2 g_{2s} (g_{1c}^2 + g_{1s}^2)
 -4 g_{1s} g_{1c} g_{20}, 
 \nonumber \\
 K_{2c} &=
 2 g_{20} (g_{1s}^2-g_{1c}^2) + 2 g_{2c} (g_{1s}^2+g_{1c}^2),
 \nonumber \\
 K_{4s} &=
 g_{2s}(g_{1c}^2-g_{1s}^2) + 2 g_{1c} g_{1s} g_{2c} - 16 g_0 g_{2c} g_{2s},
\nonumber \\
K_{4c} &=
g_{2c}(g_{1c}^2-g_{1s}^2) + 8 g_0 (g_{2s}^2-g_{2c}^2)
-2 g_{1s}g_{1c}g_{2s}.
 \nonumber
\end{align}
We now convert (\ref{eq:K}) to an equation for the roots of a polynomial, since all roots of such equations can be computed robustly using the companion matrix method described below. 
Introducing $w = \sin 2\vartheta$, then $\cos 2\vartheta = \varpi \sqrt{1-w^2}$ where $\varpi = \pm 1$. Also, $\sin 4\vartheta = 2 w \varpi \sqrt{1-w^2}$ and $\cos 4\vartheta = 1-2 w^2$. Using these results to eliminate the trigonometric functions, (\ref{eq:K}) can be written 
\begin{equation}
\label{eq:w}
    P_4 w^4 + P_3 w^3 + P_2 w^2 + P_1 w + P_0 = 0
\end{equation}
where
\begin{align}
P_4 &= 4 K_{4c}^2 + 4 K_{4s}^2,\\
P_3 &= 4 K_{4s} K_{2c} - 4 K_{4c} K_{2s},
\nonumber \\
P_2 &= K_{2s}^2+K_{2c}^2 - 4 K_0 K_{4c} - 4 K_{4c}^2 - 4 K_{4s}^2,
\nonumber \\
P_1 &= 2 K_0 K_{2s} +2 K_{4c} K_{2s} - 4 K_{4s} K_{2c},
\nonumber \\
P_0 &= (K_0 + K_{4c})^2 - K_{2c}^2. \nonumber
\end{align}
The solutions $w$ of (\ref{eq:w}) can be computed robustly by finding the eigenvalues of the companion matrix
\begin{equation}
M = \begin{pmatrix}
-P_3/P_4 & -P_2/P_4 & -P_1/P_4 & -P_0/P_4 \\
1 & 0 & 0 & 0 \\
0 & 1 & 0 & 0 \\
0 & 0 & 1 & 0
\end{pmatrix},
\end{equation}
\changed{a matrix constructed such that its characteristic polynomial $\det|M - wI|$ (for identity matrix $I$) is proportional to the original polynomial (\ref{eq:w}).}

Then for each real solution $w$,  $\varpi$ 
 can in principle be computed from (\ref{eq:K}) in the form
\begin{equation}
    \varpi = -\frac{K_0 + K_{2s} w + K_{4c}(1-2w^2)}{(2 w K_{4s}+K_{2c}) \sqrt{1-w^2}},
\label{eq:varpi_solution}
\end{equation}
where the square of the right-hand side is 1 due to (\ref{eq:w}).
Thus, precisely one of the two choices for $\varpi$ is consistent with each real root $w$.
Due to possible precision loss in the expression (\ref{eq:varpi_solution}) in finite-precision arithmetic, it is more convenient in practice to select $\varpi$
as the element of $\{-1,1\}$ that minimizes the residual in
(\ref{eq:K}).
With $\cos 2\vartheta = \varpi\sqrt{1-w^2}$ now known,  $\cos\vartheta$ can be computed from $\varsigma \sqrt{(1+\cos 2\vartheta)/2}$, with $\varsigma=\pm 1$. 
Note that $\sin\vartheta$ can be computed from $(\sin 2\vartheta)/(2 \cos\vartheta)$. The two choices of $\varsigma$ correspond to a pair of solutions of $\left\{ \sqrt{g}=0, \,  \partial\sqrt{g}/\partial\vartheta=0\right\}$ in which $r$ differs by a factor of $-1$, i.e. the two choices represent two representations $(\vartheta_0, \, r_0)$ and $(\vartheta_0+\pi, \, -r_0)$ of the same physical point. It is no loss of generality to consider only the solution with positive $r$. One can compute $r$ from (\ref{eq:r}), or from $\sqrt{g}=0$ if the denominator of (\ref{eq:r}) vanishes.
The smallest positive solution for $r$ derived from the various roots of (\ref{eq:w}) is then $\hat{r}_c$. 

By this method, the nonlinear equations for $\hat{r}_c$
can be solved without any need for an initial guess.
The resulting value of $\hat{r}_c$ can either be used directly as a good approximation for the exact solution (by which we mean a solution of $\left\{ \sqrt{g}=0, \,  \partial\sqrt{g}/\partial\vartheta=0\right\}$ in which $g_4$ and $g_5$ are retained in $\sqrt{g}$),
or it can be used as an initial guess for Newton iteration
to find the exact solution.


\subsection{Example}
\label{sec:singularityExample}

\begin{figure}
  \centering
  \includegraphics[width=3.5in]{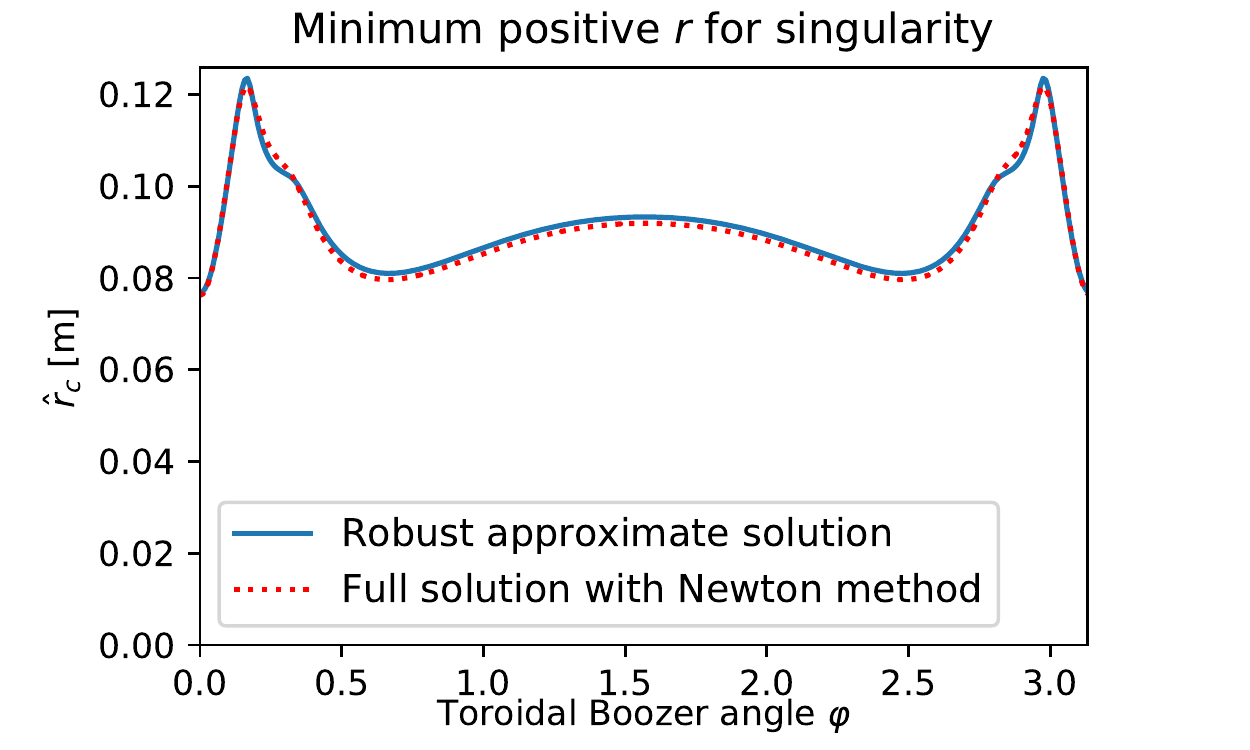}
  \caption{
Effective minor radius $r$ at which the magnetic surfaces intersect, for the example of section \ref{sec:singularityExample} and figure \ref{fig:singularity}. The blue solid curve is computed by the method of section \ref{sec:robustSingularity}.
}
\label{fig:singularityRobustVsNewton}
\end{figure}

Figures (\ref{fig:singularity})-(\ref{fig:singularityRobustVsNewton}) demonstrate
the methods of this section for an example quasi-axisymmetric configuration.
The configuration is constructed using the $O(\epsilon^2)$ equations,
with axis shape $R [\mathrm{m}]= 1 - 0.12 \cos(2\phi)$ and $Z [\mathrm{m}]=  0.12 \sin(2\phi)$. The other input parameters form the construction were $\etabar = -0.7$ m$^{-1}$, $\sigma(0)=0$, $I_2=0$, $p_2=0$, $B_{2s}=0$, and $B_{2c} = -0.5$ T$/$m$^{2}$. This configuration was chosen since multiple types of singularities can be seen in the $\phi=0$ cross-section, as shown in figure  (\ref{fig:singularity}). On the large-$R$ side, each magnetic surface with sufficiently large $r$ crosses through itself, and on the small-$R$ side, the surfaces are not simply nested beyond a critical $r$.

Figure (\ref{fig:singularityRobustVsNewton}) shows the 
value of $\hat{r}_c$ computed using the method of section \ref{sec:robustSingularity} for this example. The solution found is the one at the small-$R$ side, occurring at $\hat{r}_c=0.0762$ m in the $\varphi=0$ plane, a slightly smaller value of $r$ than the singularity at the large-$R$ side.
At each grid point in $\varphi$, this solution is used as an initial guess for a solution of the full problem from section \ref{sec:singularityProblem} using Newton's method. Newton's method converges to machine precision at all grid points, and the solution is plotted on the same graph. In this case the Newton solution barely differs from the robust approximate solution, with a solution $\hat{r}_c=0.0767$ m in the $\varphi=0$ plane. For other magnetic configurations in which the minimum aspect ratio is much lower, significant differences can arise between the robust and Newton solutions for $r_c$. 
Nevertheless, the robust method is found to be extremely accurate (as in the example shown) when the minimum aspect ratio is large, making it useful for screening out such configurations.


\subsection{Tokamak equilibrium limit}
\label{sec:betaLimit}



In both tokamaks and stellarators, an equilibrium beta limit exists associated with an X-point approaching and reaching the plasma boundary \citep{Mukhovatov, FreidbergMHDNew}.
As the method of section \ref{sec:singularityProblem}-\ref{sec:robustSingularity} finds singularities in the flux surface geometry, we might expect that this method can calculate this  equilibrium beta limit.
Here we show that this is indeed true for a
high-aspect-ratio tokamak.
This correspondence gives added meaning and value to the figure of merit $r_c$ in sections \ref{sec:singularityProblem}-\ref{sec:robustSingularity}.

For this calculation we use the ``high-$\beta$ tokamak'' ordering, eq (6.92) of \cite{FreidbergMHDNew}: $\epsilon \ll 1$ where $\epsilon = a/R$, $\mu_0 p_0 / B_0^2 \sim \epsilon$, and $\iota \sim 1$. 
Since axisymmetry is considered here, the Garren-Boozer equations in this case are given in appendix \ref{apx:axisymm}.
We consider a pressure profile $p(r) = p_0 + r^2 p_2$ such that $p(a)=0$ at the plasma boundary $r=a$, so $p_2 = -p_0 / a^2$. Then using (\ref{eq:iota_axisymm}), it follows that
\begin{align}
    \mu_0 |p_2| \gg I_2^2.
    \label{eq:betaLimitOrdering}
\end{align}

We limit our attention to up-down symmetric geometry, so $\sigma=0$. 
\changed{
The near-axis solutions have a degree of freedom $B_{2c}$ reflecting the freedom in triangularity, and for simplicity we will limit attention to the case of zero triangularity, which can be expressed as}
\begin{align}
\label{eq:noTriangularity}
    \left( X - \Delta \right)^2 + Y^2 b = c,
\end{align}
\changed{for some flux functions $\Delta(r)$, $b(r)$, and $c(r)$. This equation is independent of the near-axis expansion. To demand consistency between (\ref{eq:noTriangularity}) and the near-axis equations, we take $\Delta(r)$, $b(r)$, and $c(r)$ to be polynomials and equate terms at each order.} The $r^2 \cos^2\vartheta$ terms give the leading (constant) term of $b$ to be $X_{1c}^2/Y_{1s}^2$. Then the $r^3 \cos 3\vartheta$ terms give
\begin{align}
    Y_{1s} X_{2c} - X_{1c} Y_{2s} = 0.
\end{align}
This result, with 
(\ref{eq:X2c_axisymmetry}) and (\ref{eq:Y2c_axisymmetry}),
determines $B_{2c}$.

Using these results and the equations of appendix \ref{apx:axisymm}, and considering the major radius of the magnetic axis to be $R_0$, the first few terms in the expansion of $\sqrt{g}$ are $g_0=s_G s_\psi R_0$, $g_{1s}=0$, $g_{1c}=-2 s_G s_\psi \etabar R_0$, 
\begin{align}
    g_{20} \approx \frac{s_G s_\psi \mu_0^2 p_2^2 \left( 1 + \etabar^4 R_0^4\right)^4}{I_2^4 \etabar^2 R_0^3 \left(3 + \etabar^4 R_0^4\right)^2},
\end{align}
$g_{2s}=0$, and $g_{2c} \approx -2 g_{20}$,
where (\ref{eq:betaLimitOrdering}) has been applied.
The largest terms in (\ref{eq:K})-(\ref{eq:K0}) are then $K_0 \approx 8 g_0 g_{2c}^2$ and $K_{4c} \approx -K_0$, so $K_0 (1 - \cos 4\vartheta) = 0$, implying $\vartheta = j \pi / 2$ for integers $j$. The solutions for odd $j$ do not yield real solutions for $r$. For the even-$j$ solutions, $\partial\sqrt{g}/ \partial\vartheta=0$ is satisfied automatically. The $\vartheta=\pi$ solution is just the $\vartheta=0$ solution with $r_c \to -r_c$, so it is sufficient to consider $\vartheta=0$. Solving $0=\sqrt{g}\propto g_0 + r_c g_{1c} + r_c^2(g_{20} + g_{2c}) $ for $r_c$ then gives the positive solution to be
\begin{align}
    r_c \approx \sqrt{\frac{g_0}{g_{20}}}
    \approx \frac{I_2^2 \etabar R_0^2 (3 + \etabar^4 R_0^4)}{\mu_0 |p_2| (1 + \etabar^4 R_0^4)^2}
    =\frac{\iota_0^2 B_0^2 (3 + \etabar^4 R_0^4)}{4 \mu_0 |p_2| \etabar^3 R_0^4},
    \label{eq:rSingAxisymm}
\end{align}
where the last equality follows from (\ref{eq:iota_axisymm}).
If all terms through $g_4$ are retained in $\sqrt{g}$ instead of just terms through $g_2$, a root of $\sqrt{g}=0$ and $\partial\sqrt{g}/\partial\vartheta$ exists at $\vartheta=0$  which is identical to (\ref{eq:rSingAxisymm}) to leading order in $I_2^2 / |\mu_0 p_2| \ll 1$. Therefore the truncation at $g_2$ used for the robust solution is a good approximation.

The plasma minor radius $a$ must be within the singularity: $a < r_c$. This inequality can be written using (\ref{eq:rSingAxisymm}) in the form of a $\beta$-limit:
\begin{align}
\label{eq:myBetaLimit}
    \frac{2 \mu_0 \bar{p}}{B_0^2} < E(\etabar R_0) \iota_0^2 \frac{a}{R_0}
    \sim \iota_0^2 \frac{a}{R_0}
    ,
\end{align}
where $\bar{p} = (2/a^2) \int_0^a pr\,dr = p_0/2$ is an average pressure, and 
\begin{align}
\label{eq:betaLimitCoefficient}
    E(\etabar R_0) = \frac{3 + \etabar^4 R_0^4}{4 \etabar^3 R_0^3}
\end{align}
is \changedd{a} coefficient of order unity that depends on the elongation $\kappa_e$.
\changed{As discussed in appendix \ref{apx:axisymm},
$\kappa_e= |Y_1(\vartheta=\pi/2) / X_1(\vartheta=0)| = |Y_{1s}/X_{1c}| =  1/(\etabar^2 R_0^2)$.} For surfaces with circular cross-section, $E\to 1$.

For comparison, the tokamak equilibrium beta limit can be found in eq (236) of \cite{Mukhovatov} or section 6.5 of \cite{FreidbergMHDNew}. These previous results can be expressed as
\begin{align}
\label{eq:previousBetaLimit}
\left< \frac{2 \mu_0 p}{B^2}\right> \lesssim \iota^2 \frac{a}{R_0},    
\end{align}
where $\left< \right>$ represents an average over the plasma and $\iota$ is a characteristic rotational transform. The correspondence between (\ref{eq:myBetaLimit}) and (\ref{eq:previousBetaLimit}) is apparent. An exact match of the coefficient (\ref{eq:betaLimitCoefficient}) to earlier results like \cite{FreidbergMHDNew} is complicated because the earlier calculations include radially varying elongation and magnetic shear, which do not appear until higher orders in the Garren-Boozer expansion than the order we consider. 
The boundary separatrix in the earlier tokamak calculations cannot truly be represented by the shape (\ref{eq:radial_expansion}) and (\ref{eq:poloidal_expansions}) through $O(r^2)$ used here, so we would not necessarily expect the numerical coefficient to match exactly.
Nevertheless the basic scaling $\beta \sim \iota^2 a/R$ is in clear agreement.

Based on this successful comparison, the singularity calculation in sections \ref{sec:singularityProblem}-\ref{sec:robustSingularity} could perhaps be used to find stellarator configurations with a suitably high equilibrium beta limit, as follows. Given a desired on-axis pressure $p_0$ and desired minor radius $a$, $p_2$ could be set equal to $-p_0/a^2$. Then other parameters of the Garren-Boozer model (axis shape, $\etabar$, etc) could be found numerically such that $r_c > a$. Any resulting configuration satisfying this inequality should have an acceptably high equilibrium beta limit (at least of the sort considered here, associated with separatrices).
It should be acknowledged that stellarators also are thought to have an equilibrium beta limit associated with increasing stochasticity of the magnetic field \citep{LoizuBetaLimit}, which is not computed by the method here.

Near the equilibrium $\beta$ limit, the shift of the geometric center of the boundary flux surfaces relative to the magnetic axis (Shafranov shift) becomes a substantial fraction of the plasma minor radius. It is noteworthy that this Shafranov shift is accurately represented in the Garren-Boozer equations, as shown in detail for axisymmetry in appendix \ref{apx:axisymm}. 
Therefore, using the Garren-Boozer equations, it may also be possible to assess the equilibrium $\beta$ limit in a general stellarator without using the $r_c$ metric of section \ref{sec:singularityProblem}-\ref{sec:robustSingularity} by instead examining the shift of the surface centroids relative to the axis.
\changed{While such an approach would only detect the kind of singularity at the left of figure \ref{fig:singularity} and not the kind at the right, it may provide more insight than the  procedure of sections \ref{sec:singularityProblem}-\ref{sec:robustSingularity}.}
This alternative approach is left for future work.


\section{Error field for the first-order construction}
\label{sec:errorField}


Not only can the near-axis analysis be used to 
construct boundary shapes that give quasisymmetry to a certain order;
the expansions can also provide formulae for the size of the quasisymmetry-breaking variation in $B$ at the next order.
If the magnitude of this symmetry-breaking is large, then
the volume of good quasisymmetry is evidently quite limited,
whereas configurations with small symmetry-breaking terms
can be expected to have a larger volume of good quasisymmetry.
Therefore, we expect it will be useful to derive formulae for
these symmetry-breaking terms.

In this section, we will consider the $O(\epsilon^1)$ quasisymmetry
construction, in which the cross-section of the boundary flux surface
(in the plane perpendicular to the magnetic axis) is a perfect ellipse. We will then compute the $O(\epsilon^2)$ variation in $B$, which will
generally break the quasisymmetry. The analysis will closely
follow 
\changed{
the method of section 3 and appendix B of \LS~for computing the field strength that is realized inside a constructed boundary.
However in \LS, the surface shape coefficients $(X_j, Y_j, Z_j)$ were constructed through $j=2$, and some $j=3$ terms were included, whereas here the boundary shape includes only $j=1$ terms.
The calculation in \LS~showed what the $j=2$ and $j=3$ terms must be to achieve a desired $B_2$, whereas the calculation here shows what $B_2$ is achieved if the $j \ge 2$ terms in the boundary shape are not included. These two calculations, while related, are sufficiently different that the latter requires the separate derivation here.
}


\subsection{Expansion}

We now review the expansion developed in section 3 and appendix B of \LS.
Throughout section \ref{sec:errorField} we do not assume quasisymmetry.
We suppose a solution of the Garren-Boozer equations for $X_1$ and $Y_1$ is fixed. From this solution, a finite-aspect-ratio surface is constructed by setting $r$ to a finite value $a$. The higher-order terms $X_j$, $Y_j$, and $Z_j$ for $j > 1$ are not considered when generating this surface. Then, $\vect{J}\times\vect{B}=\nabla p$ is solved inside this boundary without making a high-aspect-ratio expansion, with the boundary condition that $\vect{B}$ has no component normal to the boundary surface. This step could be done by running the VMEC code in fixed-boundary mode using the constructed surface. The result is a configuration that is similar to but not identical to the original near-axis solution: we integrated outward from the axis only approximately, but then solved for the equilibrium inside the boundary surface exactly. As a result the final axis shape will be slightly different than the initial one. The resulting finite-aspect-ratio configuration also satisfies the Garren-Boozer equations, but with slightly altered expansion coefficients, denoted with tildes, e.g. $\tilde{X}_2$.
The original `non-tilde' expansion represents a single configuration we would like to achieve, whereas the tilde expansion represents a family of configurations parameterized by $a$. Thus, while the position vector in the non-tilde configuration is given by (\ref{eq:positionVector}), the position vector in a tilde configuration is
\begin{align}
\label{eq:positionVector_tilde0}
\tilde{\vect{r}}=
\tilde{\vect{r}}_0(a,\tilde\varphi)
+\tilde{X}(a,r,\tilde\vartheta,\tilde\varphi) \tilde{\vect{n}}(a,\tilde\varphi)
+\tilde{Y}(a,r,\tilde\vartheta,\tilde\varphi) \tilde{\vect{b}}(a,\tilde\varphi)
+\tilde{Z}(a,r,\tilde\vartheta,\tilde\varphi) \tilde{\vect{t}}(a,\tilde\varphi).
\end{align}
The Boozer angles $\tilde\vartheta$ and $\tilde\varphi$ in the constructed
configuration are allowed to differ slightly from the original Boozer angles $\vartheta$ and $\varphi$, and the differences are given by single-valued quantities $t$ and $f$:
\begin{align}
\tilde\vartheta(a,\vartheta,\varphi) = \vartheta + t(a,\vartheta,\varphi),
\hspace{0.5in}
\tilde\varphi(a,\vartheta,\varphi) = \varphi + f(a,\vartheta,\varphi).
\end{align}
These expressions define $(\tilde\vartheta,\tilde\varphi)$ only up to an additive constant, and it is convenient to eliminate this freedom using the following constraints:
\begin{align}
\label{eq:angle_average_0}
\int_0^{2\pi}d\vartheta \int_0^{2\pi}d\varphi\; t(a,\vartheta,\varphi) = 0,
\hspace{0.5in}
\int_0^{2\pi}d\vartheta \int_0^{2\pi}d\varphi\; p(a,\vartheta,\varphi) = 0.
\end{align}

Similar to (\ref{eq:radial_expansion}), the $\tilde{X}$, $\tilde{Y}$ and $\tilde{Z}$ coefficients each have expansions of the form
\begin{align}
\tilde{X}(a,r,\tilde\vartheta,\tilde\varphi) = \sum_{j=1}^{\infty} r^j \tilde{X}_j(a,\tilde\vartheta,\tilde\varphi),
\end{align}
while $\tilde{B}$ and $\tilde{\beta}$ have analogous expansions that include a $j=0$ term, e.g.
\begin{align}
\tilde{B}(a,r,\tilde\vartheta,\tilde\varphi) = \sum_{j=0}^{\infty} r^j \tilde{B}_j(a,\tilde\vartheta,\tilde\varphi).
\end{align}
For each quantity in the tilde configurations,
the $a$ dependence has a Taylor series denoted by superscripts in parentheses:
\begin{align}
\tilde{\vect{r}}_0(a,\tilde\varphi) = \sum_{k=0}^{\infty} a^k \tilde{\vect{r}}_0^{(k)}(\tilde\varphi).
\end{align}
Analogous expansions exist for $\tilde{\vect{n}}$, $\tilde{\vect{b}}$, and $\tilde{\vect{t}}$.
Similarly, 
\begin{align}
\tilde{X}_j(a,\tilde\vartheta,\tilde\varphi) = \sum_{k=0}^{\infty} a^k \tilde{X}_j^{(k)}(\tilde\vartheta,\tilde\varphi),
\end{align}
and analogous expansions exist for $\tilde{Y}_j$, $\tilde{Z}_j$, $\tilde{B}_j$, and $\tilde{\beta}_j$. The analogous expansion for the angle differences is
\begin{align}
t(a,\vartheta,\varphi) = \sum_{k=0}^{\infty} a^k t^{(k)}(\vartheta,\varphi),
&\hspace{0.5in}
f(a,\vartheta,\varphi) = \sum_{k=0}^{\infty} a^k f^{(k)}(\vartheta,\varphi).
\end{align}
The profiles $I(r)$ and $p(r)$ are
considered to be the same in the tilde and non-tilde configurations, since these profiles are typically inputs to an MHD equilibrium calculation. Therefore in the finite-minor-radius calculation they can be matched exactly to the non-tilde profiles. However, the profiles $G(r)$ and $\iota(r)$ may generally differ in the tilde configurations, so we expand
\begin{align}
\tilde{G}(a,r) = \sum_{j=0}^{\infty} r^{2j} \tilde{G}_{2j}(a),
\hspace{0.5in}
\tilde{\iota}(a,r) = \sum_{j=0}^{\infty} r^{2j} \tilde{\iota}_{2j}(a),
\end{align}
where
\begin{align}
\tilde{G}_j(a) = \sum_{k=0}^{\infty} a^{k} \tilde{G}_{j}^{(k)},
&\hspace{0.5in}
\tilde{\iota}_j(a) = \sum_{k=0}^{\infty} a^{k} \tilde{\iota}_{j}^{(a)}.
\end{align}
We emphasize that subscripts refer to an expansion in distance from the axis of a given configuration, while superscripts indicate a separate expansion in the value of minor radius substituted into the original non-tilde expansion.

At $r=a$, the boundary shape represented by the non-tilde and tilde expansions coincides exactly, since this equivalence defines the tilde configuration.
The equation representing this equivalence is,
using (\ref{eq:positionVector}) and (\ref{eq:positionVector_tilde0}), 
\begin{align}
\label{eq:positionVector_tilde}
& \vect{r}_0(\varphi)
+X(a,\vartheta,\varphi) \vect{n}(\varphi)
+Y(a,\vartheta,\varphi) \vect{b}(\varphi)
+Z(a,\vartheta,\varphi) \vect{t}(\varphi)
\\
&= 
\tilde{\vect{r}}_0(a,\tilde\varphi)
+\tilde{X}(a,a,\tilde\vartheta,\tilde\varphi) \tilde{\vect{n}}(a,\tilde\varphi)
+\tilde{Y}(a,a,\tilde\vartheta,\tilde\varphi) \tilde{\vect{b}}(a,\tilde\varphi)
+\tilde{Z}(a,a,\tilde\vartheta,\tilde\varphi) \tilde{\vect{t}}(a,\tilde\varphi),
\nonumber
\end{align}
and it  plays a central role in the analysis.

To the order of interest, the field strength in the constructed (i.e. tilde) configurations is
\begin{align}
\label{eq:constructedB}
\tilde{B}(a,r,\tilde\vartheta,\tilde\varphi)
= &\tilde{B}_0^{(0)}(\tilde\varphi)
+r \tilde{B}_1^{(0)}(\tilde\vartheta,\tilde\varphi)
+a \tilde{B}_0^{(1)}(\tilde\varphi) \\
&
+r^2 \tilde{B}_2^{(0)}(\tilde\vartheta,\tilde\varphi)
+r a \tilde{B}_1^{(1)}(\tilde\vartheta,\tilde\varphi)
+a^2 \tilde{B}_0^{(2)}(\tilde\varphi)
+O(\epsilon^3). \nonumber
\end{align}
In Appendix B.2 of \LS, it was shown that for the first-order construction,
$\tilde{B}_0^{(0)}(\varphi)=B_0(\varphi)$,
$\tilde{B}_1^{(0)}(\vartheta,\varphi)=B_1(\vartheta,\varphi)$,
and $\tilde{B}_0^{(1)}(\varphi)=0$.
\changed{Computation of the remaining terms is shown in detail in appendix \ref{apx:errorfield}.}
In subsection \ref{sec:errorfield_helical} we will compute $\tilde{B}_2^{(0)}$
and $\tilde{B}_1^{(1)}$, and in subsection \ref{sec:mirror} we will compute $\tilde{B}_0^{(2)}$. This will complete the calculation
of all terms in the second-order field strength (\ref{eq:constructedB})
in terms of non-tilde quantities, which are considered known.
 

\subsection{Summary of results}
\label{sec:errorfield_summary}

We now summarize the practical consequences of 
appendix \ref{apx:errorfield}.
When a magnetic surface is constructed
to achieve a desired $B(\vartheta,\varphi)$ to $O(\epsilon^1)$,
the $O(\epsilon^2)$ ``error field'' terms in $B$ are given
by $r^2 \tilde{B}_{2}^{(0)}(\tilde{\vartheta},\tilde{\varphi}) + a^2 \tilde{B}_0^{(2)}(\tilde{\varphi})$, since $\tilde{B}_1^{(1)}$ vanishes.
To compute these terms we first evaluate $\tilde{Z}_2^{(0)}$ from (A27)-(A31) of \LS.
Then we solve a linear system
given by (A41)-(A42) (with tildes) in \LS, together with
(\ref{eq:X2sSubstitution})-(\ref{eq:Y2cSubstitution}),
with $\tilde{X}_{20}^{(0)}$ and $\tilde{Y}_{20}^{(0)}$
as the unknowns. 
Then  $\tilde{B}_{2}^{(0)}(\tilde{\vartheta},\tilde{\varphi})=\tilde{B}_{20}^{(0)}(\tilde{\varphi})+\tilde{B}_{2s}^{(0)}(\tilde{\varphi})\sin 2\tilde{\vartheta}+\tilde{B}_{2c}^{(0)}(\tilde{\varphi})\cos 2\tilde{\vartheta}$
is evaluated from (A34)-(A40) (adding tildes and $^{(0)}$ superscripts to the subscript-2 quantities).
Finally, $\tilde{B}_0^{(2)}$ is found from
(\ref{eq:B02})-(\ref{eq:f2bar}).
\changed{For the special case of quasisymmetry, 
(\ref{eq:X2sSubstitution})-(\ref{eq:Y2cSubstitution}) simplify as shown in subsection \ref{sec:errorfield_quasisymmetry}.}


\subsection{Numerical verification}

\begin{figure}
  \centering
  \includegraphics[width=5.0in]{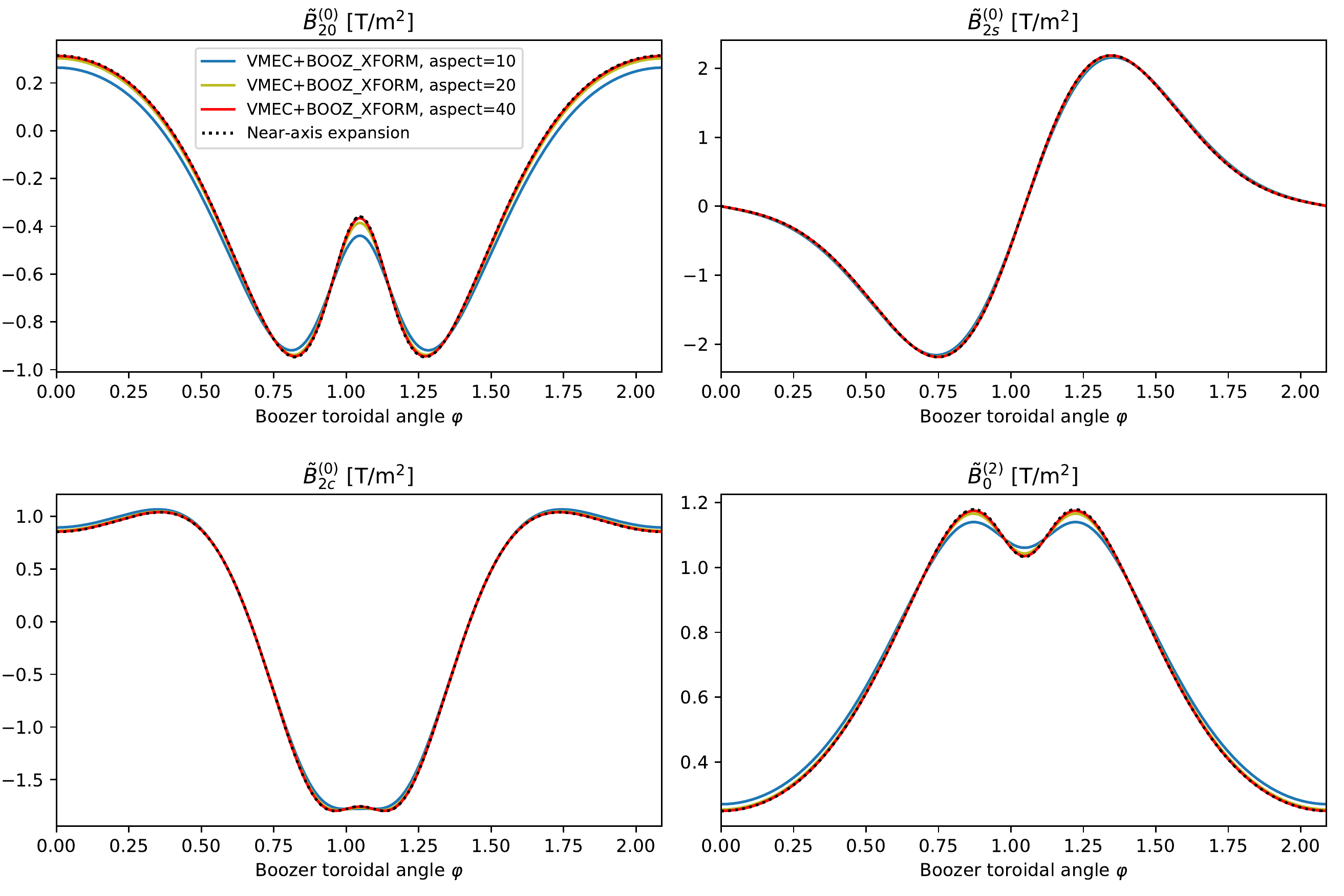}
  \caption{
The leading quasisymmetry-breaking terms in $B$ for the constructed quasi-axisymmetric example of section 5.1 of \cite{PaperII}. Excellent agreement is evident between 
the values of these terms computed by the near-axis analysis (dotted) compared to the values from finite-aspect-ratio equilibria (solid curves). The latter are computed with the VMEC and BOOZ\_XFORM codes at a range of aspect ratios.
}
\label{fig:errorField}
\end{figure}

We now demonstrate the methods of this section for
a concrete example, the quasi-axisymmetric configuration of section 5.1 in \cite{PaperII}, shown in figures 2-3 of that reference. This configuration is defined by
the axis shape $R = 1 + 0.045\cos(3\phi)$ and $Z = -0.045\sin(3\phi)$ in units of meters, where $(R,\phi,Z)$ are standard cylindrical coordinates, and $\etabar=-0.9$ m$^{-1}$, along with $I_2=0$ and $p_2=0$.
For the $O(\epsilon)$ construction, the leading-order terms in $B$ that cause departures from quasisymmetry are $\tilde{B}_{20}^{(0)}$, $\tilde{B}_{2s}^{(0)}$, $\tilde{B}_{2c}^{(0)}$, and $\tilde{B}_{0}^{(2)}$.
These four quantities are computed using the near-axis analysis of this section and plotted in figure \ref{fig:errorField} as dotted curves. Then,
toroidal magnetic surfaces are constructed for several different choices of the boundary minor radius $a$: 1/10, 1/20, and 1/40 meters, giving effective aspect ratios of 10, 20, and 40. The VMEC code \citep{VMEC1983} is used to solve for the MHD equilibrium inside the constructed boundaries without making the near-axis expansion, and the VMEC results are converted to Boozer coordinates using the BOOZ\_XFORM code \citep{Sanchez}. From these results, we can identify $\tilde{B}_{2s}^{(0)}$ and $\tilde{B}_{2c}^{(0)}$ as the magnitudes of the $\sin 2\theta$ and $\cos 2\theta$ modes at the boundary (scaled by $1/a^2$), and identify 
$\tilde{B}_{0}^{(2)}$ as the magnitude of $B$ on the magnetic axis after subtracting off $B_0 = 1$ T and scaling by $1/a^2$. We can also identify $\tilde{B}_{2}^{(0)}$ as the difference between the $\theta$-independent Fourier mode at the boundary and axis, scaled by $1/a^2$. These values of 
$\tilde{B}_{20}^{(0)}$, $\tilde{B}_{2s}^{(0)}$, $\tilde{B}_{2c}^{(0)}$, and $\tilde{B}_{0}^{(2)}$
extracted from the finite-aspect-ratio VMEC solutions are displayed in figure \ref{fig:errorField} as solid curves. It can be seen that there is excellent agreement between the quantities derived from finite-aspect-ratio VMEC solutions and the near-axis analysis, and the agreement improves as $a$ decreases.

\begin{figure}
  \centering
  \includegraphics[width=5.5in]{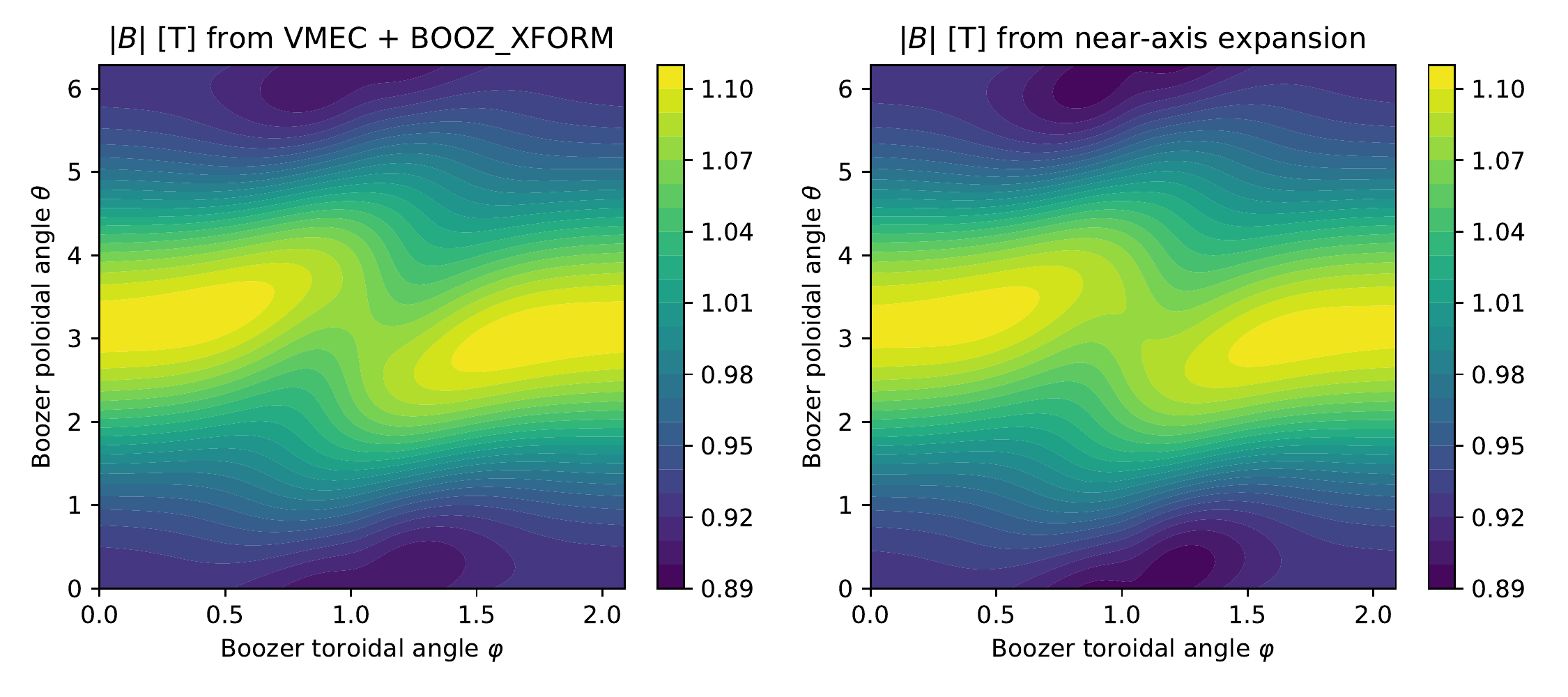}
  \caption{
Total magnetic field strength on the boundary of the constructed quasi-axisymmetric example of section 5.1 of \cite{PaperII} for aspect ratio 10.
At left is the result from a finite-aspect-ratio MHD equilibrium solution inside the constructed boundary.
At right is the result of the near-axis analysis, considering both the $O(\epsilon)$ quasisymmetric component and the $O(\epsilon^2)$ non-quasisymmetric component. The two methods evidently agree well.
}
\label{fig:totalB}
\end{figure}

Figure \ref{fig:totalB} provides another perspective on the same quasi-axisymmetric example.
The left panel shows the total field strength at the constructed $a=1/10$ boundary, as computed by finite-aspect-ratio computations with VMEC and BOOZ\_XFORM. The right panel shows the boundary field strength predicted for this case by the near-axis expansion, (\ref{eq:constructedB}), including both the $O(\epsilon)$ quasisymmetric component and the $O(\epsilon^2)$ quasisymmetry-breaking terms. Good agreement is apparent.

\section{Discussion and conclusions}

In this paper we have derived several diagnostic quantities that can be used to asses stellarator configurations. 
These diagnostics can all be computed directly from a solution of the near-axis equilibrium equations of \cite{GB2}.
The first diagnostics are the scale lengths $L_{\nabla B}$ and $L_{\nabla \nabla B}$. These quantities could be maximized, or one could require that these quantities be above some threshold, since a short scale length probably implies that an electromagnetic coil must be close to the magnetic axis. Instead, coils should be far from the magnetic axis in order to have sufficient space for a vacuum vessel and blanket, and to reduce ripple in $B$. 
The next diagnostic derived was $r_c$, the maximum minor radius at which the flux surfaces are smooth and nested. This quantity should be maximized, or one should require this quantity to be above some threshold, since otherwise a near-axis configuration may be limited to very high aspect ratio. Moreover, we showed $r_c$ is related to an equilibrium $\beta$ limit, so ensuring $r_c$ is sufficiently large at the desired pressure should ensure that this $\beta$ limit is not too severe.
The last diagnostic derived here was the field strength at $O(r^2)$ for a configuration constructed to have a desired $B$ at $O(r^1)$. This ``error field'' could be minimized 
\changed{using numerical optimization while varying the axis shape and other model parameters}
to ensure that deviations from quasisymmetry or omnigenity are not too great.

These diagnostics can all be computed by algebraic manipulation of the solution of the near-axis equilibrium equations, without requiring a  finite-aspect-ratio numerical equilibrium from a code such as VMEC. Therefore these diagnostics can all be computed within the timescale of a few milliseconds on which the near-axis equilibrium equations are solved \citep{PaperII,r2GarrenBoozer}, orders of magnitude faster than the objective function evaluations of traditional stellarator optimization. 
In future work, we intend to apply these diagnostics to both optimization and to scans over the parameter space of the near-axis equations. Due to the speed by which the near-axis equations and these diagnostics can be evaluated, it is feasible to conduct wide and high-resolution scans over parameter space. 

Another application of a result of this paper is single-stage optimization of coil shapes for quasisymmetry. This technique, explained in \cite{Giuliani}, depends fundamentally on the $\nabla\vect{B}$ tensor derived here, eq (\ref{eq:gradB_QS}). There are many possible generalization of this work, such as extending the method to $O(r^2)$ quasisymmetry using the $\nabla\nabla\vect{B}$ tensor from section \ref{sec:tensors}, and including other figures of merit in the optimization beyond those in \cite{Giuliani}.

Besides the applications mentioned above, several other avenues for future work are evident. First, 
it would be useful to verify the accuracy of 
\changedd{$L_{\nabla B}$ or $\min(L_{\nabla B}, \, L_{\nabla \nabla B})$} as surrogates for coil complexity by comparison to coil designs. For example, a database of stellarator equilibria could be gathered, and coils could be computed for each configuration using a code such as REGCOIL \citep{regcoil} or FOCUS \citep{FOCUS}. The correlation could then be checked between the complexity of these coils versus 
\changedd{$L_{\nabla B}$ or $\min(L_{\nabla B}, \, L_{\nabla \nabla B})$, depending on the order of expansion used.}
If the correlation is reasonably good, it may then be useful to compute the 
$L_{\nabla B}$ \changedd{or $\min(L_{\nabla B}, \, L_{\nabla \nabla B})$} diagnostics not only for near-axis equilibrium solutions but also for finite-aspect-ratio MHD solutions, from codes such as VMEC. One could maximize $L_{\nabla B}$ and $L_{\nabla \nabla B}$ in traditional optimization of the plasma boundary shape to ensure that the equilibria obtained do not require very close coils.
A second opportunity for future work could be to
repeat the method of section \ref{sec:errorField} at next order,
computing the $O(r^3)$ field strength for a configuration that was constructed to have a desirable $B$ through $O(r^2)$.
\changed{To extend the calculation to next order in this way, the algebra will be quite complicated, but it may be possible to make progress with assistance from symbolic algebra software.}
Finally, using the knowledge of flux surface singularities from section \ref{sec:singularity}, it may be possible to put a singularity at a desired location for a divertor.


This work was supported by the
U.S. Department of Energy, Office of Science, Office of Fusion Energy Science,
under award number DE-FG02-93ER54197.
This work was also supported by a grant from the Simons Foundation (560651, ML).
Assistance from Rogerio Jorge  with calculation of the $\nabla\nabla\vect{B}$ tensor and discussion of the $\beta$ limit are gratefully acknowledged.


\appendix

\section{Coefficients in the Jacobian}
\label{apx:JacobianCoeffs}



The $g_2$ terms in the Jacobian $\sqrt{g}$ in sections \ref{sec:singularityProblem}-\ref{sec:robustSingularity} are
\begin{align}
g_{20} = &2\kappa  \ell' X_{20} ( X_{1s} Y_{1c}- X_{1c} Y_{1s})
   +X_{2c} (-\kappa  \ell' (X_{1c} Y_{1s} + X_{1s} Y_{1c})
   +4 \ell' Y_{2s}) \nonumber \\
   &
   +X_{2s} (\kappa
    \ell' (X_{1c} Y_{1c}-X_{1s} Y_{1s})-4 \ell'
   Y_{2c})+2 \kappa  \ell' X_{1c} X_{1s} Y_{2c}+\kappa 
   \ell' Y_{2s} \left(X_{1s}^2-X_{1c}^2\right)\nonumber \\
   &
-Z_{20} (X_{1c} Y_{1s} - X_{1s} Y_{1c})'
  +Z_{20}' (X_{1c} Y_{1s}-X_{1s}
   Y_{1c})
   \nonumber \\
&    +Z_{2s} \left(-\ell' \tau V_3 +Y_{1c}
   X_{1c}'-X_{1c} Y_{1c}'-Y_{1s} X_{1s}'+X_{1s}
   Y_{1s}'\right)
   \nonumber \\
&   +Z_{2c} \left(\ell' \tau  V_2 -Y_{1s} X_{1c}'+X_{1c} Y_{1s}'-Y_{1c}
   X_{1s}'+X_{1s} Y_{1c}'\right),
\end{align}
\begin{align}
g_{2s} = & X_{20} (\kappa  \ell' (X_{1c} Y_{1c}-X_{1s} Y_{1s})-4 \ell' Y_{2c})
   +X_{2c} (-\kappa \ell' (X_{1c} Y_{1c}+X_{1s} Y_{1s})+4 \ell'
   Y_{20})\nonumber \\
   &
   +2\kappa  \ell' X_{2s} ( X_{1s} Y_{1c}-X_{1c} Y_{1s})
   +\kappa  \ell' Y_{20} \left(X_{1s}^2-X_{1c}^2\right)+\kappa  \ell' Y_{2c} \left(X_{1c}^2+X_{1s}^2\right)\nonumber \\
& + Z_{20} \left(-\ell' \tau V_3 +Y_{1c} X_{1c}'-X_{1c} Y_{1c}'-Y_{1s}
   X_{1s}'+X_{1s} Y_{1s}'\right)
   \nonumber \\
   &
-Z_{2s} (X_{1c} Y_{1s} - X_{1s} Y_{1c})'
+Z_{2s}' (X_{1c} Y_{1s}-X_{1s} Y_{1c})
   \nonumber \\
  &   +Z_{2c} \left(\ell' \tau V_1 -Y_{1c}
   X_{1c}'+X_{1c} Y_{1c}'-Y_{1s} X_{1s}'+X_{1s}
   Y_{1s}'\right)  ,
\end{align}
\begin{align}
g_{2c} = & X_{20} (-\kappa  \ell' (X_{1c} Y_{1s}+X_{1s} Y_{1c})+4 \ell' Y_{2s})
   +2\kappa  \ell' X_{2c} (X_{1s} Y_{1c}-X_{1c} Y_{1s}) \nonumber \\
   &
   +X_{2s} (\kappa
    \ell' (X_{1c} Y_{1c}+X_{1s} Y_{1s})-4 \ell'
   Y_{20})+2 \kappa  \ell' X_{1c} X_{1s} Y_{20}+\kappa 
   \ell' Y_{2s} \left(-X_{1c}^2-X_{1s}^2\right)\nonumber \\
& + Z_{20} \left(\ell' \tau  V_2 -Y_{1s} X_{1c}'+X_{1c} Y_{1s}'-Y_{1c}
   X_{1s}'+X_{1s} Y_{1c}'\right)
   \nonumber \\
   &
   +Z_{2s} \left(-\ell' \tau V_1 +Y_{1c}
   X_{1c}'-X_{1c} Y_{1c}'+Y_{1s} X_{1s}'-X_{1s}
   Y_{1s}'\right)\nonumber \\
   &
-Z_{2c} (X_{1c} Y_{1s} - X_{1s} Y_{1c})'
   +Z_{2c}' (X_{1c} Y_{1s}-X_{1s}
   Y_{1c}),
\end{align}
where, as in \cite{GB1}, $V_1 = X_{1s}^2 + X_{1c}^2 + Y_{1s}^2 + Y_{1c}^2$, $V_2 = 2 (X_{1s}X_{1c} + Y_{1s}Y_{1c})$,
and $V_3 = X_{1c}^2 - X_{1s}^2 + Y_{1c}^2 - Y_{1s}^2$.

\section{Garren-Boozer equations in axisymmetry}
\label{apx:axisymm}

In this appendix we show how Garren and Boozer's near-axis equilibrium equations reduce in the case of axisymmetry.
These results are used in section \ref{sec:betaLimit}.
Garren and Boozer's equations for general nonaxisymmetric geometry are given in the appendix of \cite{GB2} and in appendix B of \LS; here we will refer to equation numbers in the latter reference.
In axisymmetry, with the magnetic axis at major radius $R_0$, then
$\ell'=R_0$, $G_0 = s_G B_0 R_0$, $\kappa=1/R_0$, and $\tau=0$. Along the magnetic axis, $\varphi=\phi$ where $\phi$ is the standard toroidal angle. \LS~eq (A26) becomes
\begin{align}
\iota_0 = \frac{2 s_G R_0^3 \bar{\eta}^2 I_2}{B_0 (\bar{\eta}^4 R_0^4 + 1 + \sigma^2)}
\label{eq:iota_axisymm}
\end{align}
where $\sigma$ is now a constant parameter.
The shape of the flux surfaces to $O(r)$ is given by
\begin{align}
\label{eq:firstOrder_axisymmetry}
X_{1c}=\bar{\eta} R_0,
\hspace{0.3in}
Y_{1s}=\frac{s_G s_\psi}{\bar{\eta} R_0},
\hspace{0.3in}
Y_{1c}=\frac{s_G s_\psi \sigma}{\bar{\eta} R_0}.
\end{align}
The corresponding elongation $\elong(\phi)$ of the surfaces can be computed using (B.4) of \cite{PaperI}:
\begin{align}
\elong = \left[ V_1 + \sqrt{V_1^2 - 4 q^2}\right]/(2|q|),
\end{align}
where $V_1 = X_{1s}^2 + X_{1c}^2 + Y_{1s}^2 + Y_{1c}^2$ and
$q=X_{1s}Y_{1c}-X_{1c}Y_{1s} = -s_G s_\psi$.
The result in axisymmetry is
\begin{align}
\elong = \frac{1}{2}\left[
\bar{\eta}^2 R_0^2 + \frac{1}{\bar{\eta}^2 R_0^2}(1+\sigma^2)
+\sqrt{ \bar{\eta}^4 R_0^4-2+2\sigma^2 + \frac{1}{\bar{\eta}^4 R_0^4}(1+\sigma^2)^2}\right].
\end{align}
For stellarator symmetry ($\sigma=0$), this expression simplifies to $\elong=\max(\bar{\eta}^2 R_0^2, \; \bar{\eta}^{-2} R_0^{-2})$.
Under the further assumption of $\bar{\eta}=\pm 1/R_0$ so the elongation is 1, then $|\iota_0| = R_0 I_2 / B_0$, 
which matches the well-known formula for a high-aspect-ratio tokamak $|\iota| \approx (R_0/r)(B_\theta/B_0)$, with $B_{\theta}$ the poloidal field computed from Ampere's Law.

Returning to the case of general $\sigma$ and $\bar{\eta}$, 
\LS~eq (A42)
reduces to
\begin{align}
&C \iota_0 \left[
-\frac{4 \mu_0 p_2 \sigma}{B_0^2} + \bar{\eta}^2 (\sigma-4 s_G s_\psi R_0 Y_{2c})\right. \\
&\left.\hspace{1in}
-\frac{2}{B_0} \left( B_{2s} [1-3 \bar{\eta}^4 R_0^4 -3\sigma^2]+2 B_{20}\sigma+4 B_{2c}\sigma\right)\right]=0,
\nonumber
\end{align}
where $C$ is a nonzero constant. Therefore, as long as $\iota_0	 \ne 0$,
\begin{align}
\label{eq:Y2c_axisymmetry}
 Y_{2c}
=
s_G s_\psi \left[
-\frac{\mu_0 p_2 \sigma}{B_0^2  \bar{\eta}^2 R_0} +  \frac{\sigma}{4   R_0}
-\frac{B_{2s} [1-3 \bar{\eta}^4 R_0^4 -3\sigma^2]+2 B_{20}\sigma+4 B_{2c}\sigma}{2B_0  \bar{\eta}^2 R_0}
\right]
\end{align}
Then 
\LS~eq (A41)
gives, upon substitution of (\ref{eq:Y2c_axisymmetry}),
\begin{align}
\label{eq:B20_axisymmetry}
\frac{B_{20}}{B_0} = &-\frac{\mu_0 p_2}{B_0^2} +\frac{1}{3 - \bar{\eta}^4 R_0^4 +3\sigma^2} \left[
-\frac{\mu_0 p_2 F^2}{2 I_2^2 R_0^2}
+3(\bar{\eta}^4 R_0^4 - 1 - 3\sigma^2) \frac{B_{2c}}{B_0} \right. \\
&\left.
+6 \sigma (\bar{\eta}^4 R_0^4 + \sigma^2) \frac{B_{2s}}{B_0}
+\frac{\bar{\eta}^2}{2}(7-2 \bar{\eta}^4 R_0^4+4\sigma^2)
+\frac{4 I_2^2 \bar{\eta}^6 R_0^6(F-3)}{B_0^2 F^2}
 \right],
 \nonumber
\end{align}
where $F=\bar{\eta}^4 R_0^4 + \sigma^2 + 1$.
The rest of the coefficients needed to describe the flux surface shapes to $O(r^2)$ are $Z_{20}=0$,
\begin{align}
\label{eq:Z2_axisymmetry}
Z_{2s}=& \frac{s_G I_2 (F-2)}
{2 B_0 F},
\hspace{0.5in}
Z_{2c}=-\frac{s_G I_2 \sigma}{B_0 F},
\\
X_{20}=& \frac{I_2^2 \bar{\eta}^2 R_0^3}{B_0^2 F}
-\frac{\bar{\eta}^2 R_0}{2} + \frac{\mu_0 p_2 R_0}{B_0^2} + \frac{B_{20}}{B_0}R_0,
\\
X_{2s}=&\frac{2 I_2^2 \bar{\eta}^2 R_0^3 \sigma}{B_0^2 F^2}
+\frac{B_{2s}}{B_0} R_0,
\\
X_{2c}=&\frac{I_2^2 \bar{\eta}^2 R_0^3 (F-2)}{B_0^2 F^2} - \frac{\bar{\eta}^2 R_0}{2} + \frac{B_{2c}}{B_0} R_0,
\label{eq:X2c_axisymmetry}
\\
Y_{20} = &Y_{2c}
+\frac{s_G s_\psi}{\bar{\eta}^2 R_0} \left[ \frac{\mu_0 p_2 \sigma}{B_0^2} 
- \frac{B_{2s} + (B_{2c}-B_{20})\sigma}{B_0}\right],
\\
Y_{2s} = & -\frac{2 s_G s_\psi I_2^2 \bar{\eta}^4 R_0^5}{B_0^2 F^2}
+\frac{s_G s_\psi}{2 R_0}
-\frac{s_G s_\psi}{\bar{\eta}^2 R_0} \left[ \frac{\mu_0 p_2}{B_0^2} 
+ \frac{B_{20} + B_{2c}-B_{2s}\sigma}{B_0}\right].
\label{eq:Y2s_axisymmetry}
\end{align}
We can thus consider the equilibrium to be parameterized by the constants $R_0$,
$B_0$, $I_2$, $\bar{\eta}$, $\sigma$, $p_2$, $B_{2c}$, and $B_{2s}$; then $B_{20}$ is determined by (\ref{eq:B20_axisymmetry}), and the surface shapes
are given by (\ref{eq:firstOrder_axisymmetry}), (\ref{eq:Y2c_axisymmetry}), 
and (\ref{eq:Z2_axisymmetry})-(\ref{eq:Y2s_axisymmetry}).
The fact that $Z_2 \ne 0$ in general reflects the fact that the Boozer toroidal angle is generally only identical to the cylindrical coordinate angle $\phi$ on the axis.	

We now demonstrate the above results are consistent with the textbook result for the Shafranov shift in
a tokamak with circular cross section.
First, to consider surfaces that are circles to $O(r)$, we set $\sigma=0$
and $\bar{\eta}=1/R_0$, so $F=2$. Then to consider surfaces that remain circular to $O(r^2)$, we plug (\ref{eq:radial_expansion}) and (\ref{eq:poloidal_expansions}) (and the analogous expansion of $Y$) into the equation for a shifted circle, $(X-\Delta)^2 + Y^2 = r^2$,
assuming $\Delta = r^2 \Delta_2 + O(r^3)$.
Collecting terms with shared $\theta$ dependence, one finds $\Delta_2 = X_{20}+X_{2c}$
and $X_{2c} = s_G s_\psi Y_{2s}$.
Then (\ref{eq:B20_axisymmetry})-(\ref{eq:Y2s_axisymmetry}) give
the distance between the axis and the geometric center of the surface at radius $r$ to be
\begin{align}
\Delta = r^2 \left( \frac{1}{8 R_0} - \frac{\mu_0 p_2}{2 I_2^2 R_0}\right)
= r^2 \left( \frac{1}{8 R_0} - \frac{\mu_0 p_2 R_0}{2 \iota_0^2 B^2_{0}}\right).
\label{eq:ShafranovShiftAxisymmetry}
\end{align}
For comparison, analytic reduction of the Grad-Shafranov equation for shifted-circle
geometry gives (eq (150) in \cite{HazeltineMeiss} or eq (3.6.7) in \cite{Wesson})
\begin{align}
\frac{d}{dr}\left( r B_\theta^2 \frac{d\Delta}{dr}\right)
= \frac{r}{R_0} \left( B_\theta^2 - 2 \mu_0 r \frac{dp}{dr}\right),
\end{align}
where $B_\theta = \iota_0 B_0 r/R_0$. Substitution of $p\approx p_0 + r^2 p_2$
and $\Delta \approx r^2 \Delta_2$ then gives the same near-axis shift (\ref{eq:ShafranovShiftAxisymmetry}) as the Garren-Boozer equations.


\section{Details of the error field calculation}
\label{apx:errorfield}

\changed{
In this appendix, details are given of the calculation leading to the result in section \ref{sec:errorfield_summary}.
}
We first note that by examining (\ref{eq:positionVector_tilde}) in Appendix B.2 of \LS, it was shown that for the first-order construction,
$f^{(0)}=0$, $f^{(1)}=0$, $t^{(0)}=0$, 
$\tilde{\vect{r}}_0^{(0)}(\varphi)=\vect{r}_0(\varphi)$,
$\tilde{\vect{t}}^{(0)}(\varphi)=\vect{t}(\varphi)$,
$\tilde{\vect{n}}^{(0)}(\varphi)=\vect{n}(\varphi)$,
$\tilde{\vect{b}}^{(0)}(\varphi)=\vect{b}(\varphi)$,
$\tilde{\vect{r}}_0^{(1)}=0$,
$\tilde{\vect{t}}^{(1)}=0$,
$\tilde{\vect{n}}^{(1)}=0$,
$\tilde{\vect{b}}^{(1)}=0$,
$\tilde{X}_1^{(0)}(\vartheta,\varphi)=X_1(\vartheta,\varphi)$,
$\tilde{Y}_1^{(0)}(\vartheta,\varphi)=Y_1(\vartheta,\varphi)$.
$\tilde{G}_0^{(0)}=G_0$, $\tilde{G}_0^{(1)}=0$,
and $\tilde{\iota}_0^{(0)}=\iota_0$.

\subsection{Helical modes}
\label{sec:errorfield_helical}

We begin with the $O(\epsilon^2)$ terms in (\ref{eq:positionVector_tilde}), eq (B12) in \LS, copied here:
\begin{align}
\label{eq:a2}
&X_2(\vartheta,\varphi) \vect{n} + Y_2(\vartheta,\varphi)\vect{b}+ Z_2(\vartheta,\varphi)\vect{t} \\
&= \tilde{\vect{r}}_0^{(2)}
+f^{(1)} \tilde{\vect{r}}_0^{(1)\prime}
+f^{(2)}(\vartheta,\varphi) \vect{r}'_0
+\frac{1}{2} f^{(1)2} \vect{r}''_0 \nonumber\\
&+\tilde{X}_2^{(0)}(\vartheta,\varphi)\vect{n}
+\tilde{X}_1^{(1)}(\vartheta,\varphi)\vect{n}
+X_1(\vartheta,\varphi)\tilde{\vect{n}}^{(1)}
+f^{(1)}X_1(\vartheta,\varphi)\vect{n}'
\nonumber \\
&+t^{(1)}(\vartheta,\varphi) \vect{n} \partial_1 X_1(\vartheta,\varphi)
+f^{(1)} \vect{n} \partial_2 X_1(\vartheta,\varphi)
\nonumber \\
&+\tilde{Y}_2^{(0)}(\vartheta,\varphi)\vect{b}
+\tilde{Y}_1^{(1)}(\vartheta,\varphi)\vect{b}
+Y_1(\vartheta,\varphi)\tilde{\vect{b}}^{(1)}
+f^{(1)}Y_1(\vartheta,\varphi)\vect{b}'
\nonumber \\
&+t^{(1)}(\vartheta,\varphi) \vect{b} \partial_1 Y_1(\vartheta,\varphi)
+f^{(1)} \vect{b} \partial_2 Y_1(\vartheta,\varphi)
+\tilde{Z}_2^{(0)}(\vartheta,\varphi)\vect{t}
\nonumber 
,
\end{align}
where $\partial_1$ and $\partial_2$ indicate partial derivatives with respect to the first or second argument.
To simplify notation, the argument is not shown
for functions of only $\varphi$.
We assume the construction is done only through $O(\epsilon^1)$, so the left-hand side of (\ref{eq:a2}) is zero.
As mentioned above, in appendix B of \LS, it was proved that 
$\tilde{\vect{r}}_0^{(1)}=0$,
$\tilde{\vect{n}}^{(1)}=0$,
$\tilde{\vect{b}}^{(1)}=0$,
and
$f^{(1)}=0$ for the first-order construction.
Then (\ref{eq:a2}) reduces to
\begin{align}
\label{eq:a2reduced}
0&= \tilde{\vect{r}}_0^{(2)}
+f^{(2)}(\vartheta,\varphi) \vect{r}'_0
+\tilde{X}_2^{(0)}(\vartheta,\varphi)\vect{n}
+\tilde{X}_1^{(1)}(\vartheta,\varphi)\vect{n}
+t^{(1)}(\vartheta,\varphi) \vect{n} \partial_1 X_1(\vartheta,\varphi)
 \\
&+\tilde{Y}_2^{(0)}(\vartheta,\varphi)\vect{b}
+\tilde{Y}_1^{(1)}(\vartheta,\varphi)\vect{b}
+t^{(1)}(\vartheta,\varphi) \vect{b} \partial_1 Y_1(\vartheta,\varphi)
+\tilde{Z}_2^{(0)}(\vartheta,\varphi)\vect{t}
\nonumber 
.
\end{align}
The $\vect{t}$ component is
\begin{equation}
\label{eq:a2tangent}
0= \vect{t}\cdot\tilde{\vect{r}}_0^{(2)}
+\ell' f^{(2)}(\vartheta,\varphi) 
+\tilde{Z}_2^{(0)}(\vartheta,\varphi).
\end{equation}
The last quantity, $\tilde{Z}_2^{(0)}(\vartheta,\varphi)$,
is known since it is a unique function of $X_1$ and $Y_1$,
eq (A27)-(A29) of \LS. Therefore

\begin{equation}
\label{eq:f2}
f^{(2)}(\vartheta,\varphi) 
= \bar{f}^{(2)}(\varphi)
-\left[ \tilde{Z}_{2s}^{(0)}(\varphi) \sin 2\vartheta
+\tilde{Z}_{2c}^{(0)}(\varphi) \cos 2\vartheta\right] / \ell',
\end{equation}
where $\bar{f}^{(2)}(\varphi)$ has not yet been determined.

The argument in the paragraph of \LS~containing (B24)-(B26)
applies here, except with $X_{2s}=X_{2c}=Y_{2s}=Y_{2c}=0$, so
\begin{equation}
\label{eq:t1X}
t^{(1)}(\vartheta,\varphi) = \bar{t}^{(1)}
+\frac{2 }{X_{1s}^2 + X_{1c}^2}\left[
\left(X_{1c} \tilde{X}_{2s}^{(0)} - X_{1s} \tilde{X}_{2c}^{(0)}\right)
\cos\vartheta
-\left(X_{1s} \tilde{X}_{2s}^{(0)} + X_{1c} \tilde{X}_{2c}^{(0)}\right)
\sin\vartheta \right]
\end{equation}
and
\begin{equation}
\label{eq:t1Y}
t^{(1)}(\vartheta,\varphi) = \bar{t}^{(1)}
+\frac{2 }{Y_{1s}^2 + Y_{1c}^2}\left[
\left(Y_{1c} \tilde{Y}_{2s}^{(0)} - Y_{1s} \tilde{Y}_{2c}^{(0)}\right)
\cos\vartheta
-\left(Y_{1s} \tilde{Y}_{2s}^{(0)} + Y_{1c} \tilde{Y}_{2c}^{(0)}\right)
\sin\vartheta \right],
\end{equation}
where $\bar{t}^{(1)}(\varphi)$ has yet to be determined.
Equating the $\sin\vartheta$ and $\cos\vartheta$ modes of 
(\ref{eq:t1X}) and (\ref{eq:t1Y}), we find
\begin{align}
\label{eq:newB27}
\frac{\tilde{X}_{2s}^{(0)} X_{1s}+\tilde{X}_{2c}^{(0)} X_{1c} }{X_{1s}^2 + X_{1c}^2}
=&
\frac{\tilde{Y}_{2s}^{(0)} Y_{1s}+\tilde{Y}_{2c}^{(0)} Y_{1c} }{Y_{1s}^2 + Y_{1c}^2},\\
\label{eq:newB28}
\frac{\tilde{X}_{2c}^{(0)} X_{1s}-\tilde{X}_{2s}^{(0)} X_{1c} }{X_{1s}^2 + X_{1c}^2}
=&
\frac{\tilde{Y}_{2c}^{(0)} Y_{1s}-\tilde{Y}_{2s}^{(0)} Y_{1c} }{Y_{1s}^2 + Y_{1c}^2}.
\end{align}
Equations (\ref{eq:newB27})-(\ref{eq:newB28}) provide two linear equations constraining the subscript-2 shape coefficients
$\tilde{X}_{20}^{(0)}$,
$\tilde{X}_{2s}^{(0)}$,
$\tilde{X}_{2c}^{(0)}$,
$\tilde{Y}_{20}^{(0)}$,
$\tilde{Y}_{2s}^{(0)}$, and
$\tilde{Y}_{2c}^{(0)}$.
These six quantities are also constrained by equations (A32), (A33), (A41), and (A42) of \LS.
\changed{(Eq (A32)-(A33) express the equality at second order of the $\vect{t}$ components of the covariant and contravariant $\vect{B}$ representations, while (A41)-(A42) are the second-order version of the relation between on-axis transform, axis torsion, rotating elongation, and toroidal current.)
Thus we have} 
six linear equations for six unknowns, so we can solve the linear system. Four of the six equations are algebraic rather than differential, so the system can be reduced to a system of two equations and two unknowns for faster numerical solution. This reduction can be accomplished by considering $\tilde{X}_{20}^{(0)}$ and $\tilde{Y}_{20}^{(0)}$ as the unknowns,
solving (\ref{eq:newB27})-(\ref{eq:newB28}) with \LS~equations (A32)-(A33) to obtain
\begin{align}
\label{eq:X2sSubstitution}
\tilde{X}_{2s}^{(0)} &= \frac{1}{2b}
\left[ (X_{1s}Y_{1s}-X_{1c}Y_{1c})\tilde{X}_{20}^{(0)}
+(X_{1c}^2-X_{1s}^2)\tilde{Y}_{20}^{(0)}
+\frac{s_G \bar{B} X_{1s}X_{1c}\kappa}{B_0}\right],
\\
\tilde{X}_{2c}^{(0)} &= \frac{1}{2b}
\left[ (X_{1s}Y_{1c}+X_{1c}Y_{1s})\tilde{X}_{20}^{(0)}
-2 X_{1c} X_{1s}\tilde{Y}_{20}^{(0)}
+\frac{s_G \bar{B} (X_{1c}^2 - X_{1s}^2)\kappa}{2B_0}\right],
\\
\tilde{Y}_{2s}^{(0)} &= \frac{1}{2b}
\left[ (Y_{1s}^2-Y_{1c}^2)\tilde{X}_{20}^{(0)}
+(X_{1c} Y_{1c}-X_{1s} Y_{1s})\tilde{Y}_{20}^{(0)}
+\frac{s_G \bar{B} (X_{1s}Y_{1c}+X_{1c}Y_{1s})\kappa}{2B_0}\right],
\\
\tilde{Y}_{2c}^{(0)} &= \frac{1}{2b}
\left[ 2 Y_{1s} Y_{1c}\tilde{X}_{20}^{(0)}
-(X_{1s} Y_{1c} + X_{1c} Y_{1s})\tilde{Y}_{20}^{(0)}
+\frac{s_G \bar{B} (X_{1c}Y_{1c}-X_{1s}Y_{1s})\kappa}{2B_0}\right],
\label{eq:Y2cSubstitution}
\end{align}
where $b=X_{1s} Y_{1c}-X_{1c} Y_{1s}=-s_G \bar{B}/B_0$.
With 
$\tilde{X}_{20}^{(0)}$,
$\tilde{X}_{2s}^{(0)}$, and
$\tilde{X}_{2c}^{(0)}$ now known,
we can evaluate the tilde version of (A34)-(A36) in \LS~\changed{(which express equality of $B^2$ to the square of the contravariant representation)} to find 
$\tilde{B}_{20}^{(0)}$, 
$\tilde{B}_{2s}^{(0)}$, and
$\tilde{B}_{2c}^{(0)}$.
We have thus determined $\tilde{B}_2^{(0)}$, the first  term on the bottom line of (\ref{eq:constructedB}).

Given (\ref{eq:t1X})-(\ref{eq:t1Y}), and noting that the $\sin\vartheta$ and $\cos\vartheta$ modes of $t^{(1)}$
do not contribute to eq (B15)-(B17) of \LS,
eq (B29) holds after substituting $t^{(1)}\to \bar{t}^{(1)}$.
Then (B30) and the argument that follows it imply 
$\bar{t}^{(1)}=0$, so 
$\tilde{X}_{1s}^{(1)} = \tilde{X}_{1c}^{(1)} = \tilde{Y}_{1s}^{(1)} = \tilde{Y}_{1c}^{(1)} = 0$,
and $\tilde{B}_{1}^{(1)}=0$.
This determines the middle term on the bottom line of (\ref{eq:constructedB}).
 
 Now that $\tilde{X}_2^{(0)}$, $\tilde{Y}_2^{(0)}$, and $t^{(1)}$ are known,
the $\vect{n}$ and $\vect{t}$ components of (\ref{eq:a2reduced}) give $\tilde{\vect{r}}_0^{(2)}\cdot\vect{n}$ and $\tilde{\vect{r}}_0^{(2)}\cdot\vect{b}$. However these components
of $\tilde{\vect{r}}_0^{(2)}$ are not needed for the remainder of the calculation here, so these results will not be displayed here.


\subsection{Mirror term}
\label{sec:mirror}


It remains to compute the quantity $\tilde{B}_0^{(2)}(\tilde{\varphi})$ in (\ref{eq:constructedB}), which represents a ``mirror term'' in the field strength. To compute this term, we proceed to
examine the $O(\epsilon^3)$ terms in (\ref{eq:positionVector_tilde}).
Compared to (B32) in \LS, there are two differences in the present context. First, $f^{(2)}$ now depends on $\vartheta$, as we found in (\ref{eq:f2}). Second, $t^{(1)}$ is nonzero, given by (\ref{eq:t1X}) or (\ref{eq:t1Y}) with $\bar{t}^{(1)}=0$. Thus, we have
\begin{align}
&X_3(\vartheta,\varphi)\vect{n} + Y_3(\vartheta,\varphi)\vect{b} + Z_3(\vartheta,\varphi)\vect{t}
=\tilde{\vect{r}}_0^{(3)} + f^{(3)}(\vartheta,\varphi) \vect{r}'_0 
+ \vect{t} \tilde{Z}_3^{(0)}(\vartheta,\varphi)
+ \vect{t} \tilde{Z}_2^{(1)}(\vartheta,\varphi) \nonumber \\
&+\vect{n} \tilde{X}_3^{(0)}(\vartheta,\varphi) + \tilde{\vect{n}}^{(2)}X_1(\vartheta,\varphi)
+\vect{n}\tilde{X}_1^{(2)}(\vartheta,\varphi)+\vect{n}\tilde{X}_2^{(1)}(\vartheta,\varphi)
\nonumber \\
&+\vect{b} \tilde{Y}_3^{(0)}(\vartheta,\varphi) + \tilde{\vect{b}}^{(2)}Y_1(\vartheta,\varphi)
+\vect{b}\tilde{Y}_1^{(2)}(\vartheta,\varphi)+\vect{b}\tilde{Y}_2^{(1)}(\vartheta,\varphi)
\nonumber \\
&+f^{(2)}(\vartheta,\varphi) Y_1(\vartheta,\varphi) \vect{b}' + f^{(2)}(\vartheta,\varphi) X_1(\vartheta,\varphi) \vect{n}' 
\nonumber \\
&+\vect{n} f^{(2)}(\vartheta,\varphi) \partial_2 X_1(\vartheta,\varphi) + \vect{b} f^{(2)}(\vartheta,\varphi) \partial_2 Y_1 (\vartheta,\varphi)
\nonumber \\
&+ t^{(2)}(\vartheta,\varphi) \vect{n} \partial_1 X_1(\vartheta,\varphi) + t^{(2)}(\vartheta,\varphi) \vect{b} \partial_1 Y_1 (\vartheta,\varphi)
\nonumber \\
&+t^{(1)}(\vartheta,\varphi) \left[
\vect{n} \partial_1 \tilde{X}_2^{(0)}(\vartheta,\varphi)
+\vect{b} \partial_1 \tilde{Y}_2^{(0)}(\vartheta,\varphi)
+\vect{t} \partial_1 \tilde{Z}_2^{(0)}(\vartheta,\varphi)
\right]
\nonumber \\
&+\frac{1}{2}\left[t^{(1)}(\vartheta,\varphi)\right]^2 \left[
\vect{n}\partial_1^2 X_1(\vartheta,\varphi)
+\vect{b} \partial_1^2 Y_1(\vartheta,\varphi) \right].
\label{eq:a3}
\end{align}
The last two rows are new compared to (B32) in \LS.
We take the $\vect{n}$ component, noting $\vect{n}\cdot\vect{n}'=0$ and $\vect{n}\cdot \tilde{\vect{n}}^{(2)}=0$,
since the latter is the $O(\epsilon^2)$ term in $|\tilde{\vect{n}}|^2=1$. The $\sin\vartheta$ and $\cos\vartheta$ modes of the result are similar to (B33) of \LS, but with a few extra terms:
\begin{align}
\label{eq:X1s2}
\tilde{X}_{1s}^{(2)} =& X_{3s1} - \tilde{X}_{3s1}^{(0)} - \vect{n}\cdot\tilde{\vect{b}}^{(2)}Y_{1s}
 - X_{1s} t_{sc}^{(2)} + X_{1c} t_{ss}^{(2)} - T_{Xs}
\\
&+\left( \frac{Y_{1c} \tilde{Z}_{2s}^{(0)} - Y_{1s} \tilde{Z}_{2c}^{(0)}}{2 \ell'} - \bar{f}^{(2)}Y_{1s}\right)
\vect{n}\cdot\vect{b}'
-\left( \frac{\tilde{Z}_{2c}^{(0)}}{2\ell'}+\bar{f}^{(2)}\right) X'_{1s}
+\frac{\tilde{Z}_{2s}^{(0)}}{2\ell'} X'_{1c} \nonumber,
\\
\label{eq:X1c2}
\tilde{X}_{1c}^{(2)} =& X_{3c1} - \tilde{X}_{3c1}^{(0)} - \vect{n}\cdot\tilde{\vect{b}}^{(2)}Y_{1c}
 - X_{1s} t_{cc}^{(2)}+ X_{1c} t_{sc}^{(2)} - T_{Xc} \\
& + \left( \frac{Y_{1c}\tilde{Z}_{2c}^{(0)}+Y_{1s}\tilde{Z}_{2s}^{(0)}}{2\ell'}-\bar{f}^{(2)} Y_{1c}\right) \vect{n}\cdot\vect{b}'
+\left(\frac{\tilde{Z}_{2c}^{(0)}}{2\ell'}-\bar{f}^{(2)}\right)X'_{1c} + \frac{\tilde{Z}_{2s}^{(0)}}{2\ell'} X'_{1s}, \nonumber
\end{align}
where (\ref{eq:f2}) has been used,
\begin{align}
\label{eq:tss}
t_{ss}^{(2)}(\varphi) =& \pi^{-1} \int_0^{2\pi}d\vartheta \, t^{(2)}(\vartheta,\varphi) \sin^2\vartheta, \\
t_{sc}^{(2)}(\varphi) =& \pi^{-1} \int_0^{2\pi}d\vartheta \, t^{(2)}(\vartheta,\varphi) \sin\vartheta \cos\vartheta, \\
t_{cc}^{(2)}(\varphi) =& \pi^{-1} \int_0^{2\pi}d\vartheta \, t^{(2)}(\vartheta,\varphi) \cos^2\vartheta,
\label{eq:tcc}
\end{align}
\begin{align}
\label{eq:TXs}
T_{Xs}=\frac{1}{\pi}\int_0^{2\pi}d\vartheta \sin\vartheta
\left(
t^{(1)}(\vartheta,\varphi) \partial_1 \tilde{X}_2^{(0)}(\vartheta,\varphi)
+\frac{1}{2}\left[ t^{(1)}(\vartheta,\varphi)\right]^2 \partial_1^2 X_1(\vartheta,\varphi) \right),
\end{align}
and $T_{Xc}$ is defined by (\ref{eq:TXs}) with $\sin \to \cos$.
Similarly, taking the $\vect{b}$ component of (\ref{eq:a3}), the $\sin\vartheta$ and $\cos\vartheta$ modes give
\begin{align}
\label{eq:Y1s2}
\tilde{Y}_{1s}^{(2)} =& Y_{3s1} - \tilde{Y}_{3s1}^{(0)} - \vect{b}\cdot\tilde{\vect{n}}^{(2)}X_{1s}
 - Y_{1s} t_{sc}^{(2)} + Y_{1c} t_{ss}^{(2)} - T_{Ys} \\
&+\left( \frac{X_{1c} \tilde{Z}_{2s}^{(0)} - X_{1s} \tilde{Z}_{2c}^{(0)}}{2 \ell'} - \bar{f}^{(2)}X_{1s}\right)
\vect{b}\cdot\vect{n}'
-\left( \frac{\tilde{Z}_{2c}^{(0)}}{2\ell'}+\bar{f}^{(2)}\right) Y'_{1s}
+\frac{\tilde{Z}_{2s}^{(0)}}{2\ell'} Y'_{1c} \nonumber,
\\
\label{eq:Y1c2}
\tilde{Y}_{1c}^{(2)} =& Y_{3c1} - \tilde{Y}_{3c1}^{(0)} - \vect{b}\cdot\tilde{\vect{n}}^{(2)}X_{1c}
 - Y_{1s} t_{cc}^{(2)} + Y_{1c} t_{sc}^{(2)} - T_{Yc} \\
&+ \left( \frac{X_{1c}\tilde{Z}_{2c}^{(0)}+X_{1s}\tilde{Z}_{2s}^{(0)}}{2\ell'}-\bar{f}^{(2)} X_{1c}\right) \vect{b}\cdot\vect{n}'
+\left(\frac{\tilde{Z}_{2c}^{(0)}}{2\ell'}-\bar{f}^{(2)}\right)Y'_{1c} + \frac{\tilde{Z}_{2s}^{(0)}}{2\ell'} Y'_{1s}, \nonumber
\end{align}
where
\begin{align}
\label{eq:TYs}
T_{Ys}=\frac{1}{\pi}\int_0^{2\pi}d\vartheta \sin\vartheta
\left(
t^{(1)}(\vartheta,\varphi) \partial_1 \tilde{Y}_2^{(0)}(\vartheta,\varphi)
+\frac{1}{2}\left[ t^{(1)}(\vartheta,\varphi)\right]^2 \partial_1^2 Y_1(\vartheta,\varphi) \right) ,
\end{align}
and $T_{Yc}$ is defined by (\ref{eq:TYs}) with $\sin \to \cos$.

We now consider the $O(\epsilon^2)$ terms of (A21) in \LS, \changed{ expressing the condition that the $O(r^1 a^2)$ flux surfaces enclose the proper toroidal flux}:
\begin{align}
X_{1c} \tilde{Y}_{1s}^{(2)} + \tilde{X}_{1c}^{(2)} Y_{1s}
-X_{1s} \tilde{Y}_{1c}^{(2)} - \tilde{X}_{1s}^{(2)} Y_{1c}
=- s_G \bar{B} \tilde{B}_0^{(2)} / B_{0}^2.
\end{align}
Substituting (\ref{eq:X1s2}), (\ref{eq:X1c2}), (\ref{eq:Y1s2}), and (\ref{eq:Y1c2}) into this expression, terms involving
$\tilde{\vect{n}}^{(2)}$, $\tilde{\vect{b}}^{(2)}$, and $t^{(2)}$ cancel to leave
\begin{align}
\label{eq:B02}
\tilde{B}_0^{(2)}=\hat{B} B_0
-\bar{f}^{(2)} B'_0,
\end{align}
where
\begin{align}
\label{eq:Bhat}
\hat{B}= \left(
X_{1s}Y_{3c1} + Y_{1c}X_{3s1}-X_{1c}Y_{3s1}-Y_{1s}X_{3c1}-\tilde{q}\right) s_G B_0 / \bar{B},
\end{align}
\begin{align}
\tilde{q} = \tilde{Q}^{(0)} + (A/2) + T,
\end{align}
\begin{align}
\label{eq:Q}
\tilde{Q}^{(0)}(\varphi) =& \frac{(G_2 + I_2 N)\bar{B} \ell'}{2G_0^2} 
  +2(\tilde{X}_{2c}^{(0)} \tilde{Y}_{2s}^{(0)} - \tilde{X}_{2s}^{(0)} \tilde{Y}_{2c}^{(0)}) 
 + \frac{\bar{B}}{2G_0} \left( \ell'  \tilde{X}_{20}^{(0)} \kappa - \tilde{Z}_{20}^{(0)\prime}\right) 
 \\
& +\frac{I_2}{4G_0} \left(  -\ell'   \tau V_1 + Y_{1c} X'_{1c} - X_{1c}Y'_{1c} +Y_{1s} X'_{1s} -X_{1s}Y'_{1s}\right)
 \nonumber \\
& +\frac{\beta_0 \bar{B}}{4G_0} \left(   X_{1s} Y'_{1c} + Y_{1c} X'_{1s} - X_{1c} Y'_{1s} - Y_{1s} X'_{1c}\right), \nonumber
\end{align}
\begin{align}
A=&-2 \tau \tilde{Z}_{2c}^{(0)}(X_{1s} X_{1c}+Y_{1s}Y_{1c})
-\tau \tilde{Z}_{2s}^{(0)}(X_{1s}^2-X_{1c}^2 - Y_{1c}^2+Y_{1s}^2) 
\\
&+\frac{\tilde{Z}_{2s}^{(0)}}{\ell'} (X_{1c}Y'_{1c}+Y_{1s}X'_{1s} - X_{1s}Y'_{1s}-Y_{1c}X'_{1c}) \nonumber \\
&-\frac{\tilde{Z}_{2c}^{(0)}}{\ell'}
(X_{1c} Y'_{1s}-Y_{1s}X'_{1c}+X_{1s}Y'_{1c}-Y_{1c}X'_{1s}), \nonumber
\end{align}
and $T=
X_{1s} T_{Yc} + Y_{1c} T_{Xs}
-X_{1c} T_{Ys}
-Y_{1s}T_{Xc} 
$.
To obtain these results we have used $\vect{b}\cdot\vect{n}'=-\vect{n}\cdot\vect{b}'=\tau \ell'$,
the tilde version of (A49) of \LS~\changed{(in which the $\vect{t}$ components of the covariant and contravariant $\vect{B}$ representations are equated at third order),}
and the $d/d\varphi$ derivative of $X_{1c}Y_{1s}-X_{1s}Y_{1c}=s_G \bar{B}/B_0$.
Using (\ref{eq:t1X}), (\ref{eq:newB27}), and (\ref{eq:newB28}) to eliminate $\tilde{Y}_2^{(0)}$, we find
\begin{align}
T = 2\left[
\left( \tilde{X}_{2s}^{(0)}\right)^2 + \left(\tilde{X}_{2c}^{(0)}\right)^2 \right]
\frac{X_{1s}Y_{1c}-X_{1c}Y_{1s}}{X_{1s}^2 + X_{1c}^2}.
\end{align}
Expressions (\ref{eq:B02}), (\ref{eq:Bhat}), and (\ref{eq:Q}) mirror (3.10)-(3.12) of \LS, where $\tilde{B}_0^{(2)}$ was computed for the $O(\epsilon^2)$ construction (in contrast to the $O(\epsilon)$ construction here.)
Just as described at the end of Appendix B of \LS, $\bar{f}^{(2)}$ is determined by
\begin{align}
\bar{f}^{(2)}(\varphi) = 
\left( \int_0^{\varphi} d\bar\varphi \, \hat{B}(\bar\varphi) \right)
+\left( \frac{1}{2} - \frac{\varphi}{2\pi}\right) \left( \int_0^{2\pi} d\bar\varphi \, \hat{B}(\bar\varphi) \right)
-\frac{1}{2\pi} \int_0^{2\pi} d\bar{\bar\varphi}  \int_0^{\bar{\bar\varphi}} d\bar\varphi \, \hat{B}(\bar\varphi) .
\label{eq:f2bar}
\end{align}

At this point we have fully determined $\tilde{B}_0^{(2)}$ in terms of known quantities.
Normally, the $O(\epsilon)$ construction would be carried out with
$X_{3s1}=X_{3c1}=Y_{3s1}=Y_{3c1}=0$, so these terms would be absent in (\ref{eq:Bhat}). Alternatively, nonzero values of $X_{3s1}$, $X_{3c1}$, $Y_{3s1}$, and $Y_{3c1}$ could be chosen in order to make $\tilde{B}_0^{(2)}$ vanish, using (3.14)-(3.15) of \LS~with $Q \to \tilde{q}$.


\subsection{Quasisymmetry}
\label{sec:errorfield_quasisymmetry}

In quasisymmetry,  $X_{1s}=0$ and $\bar{B}=s_\psi B_0$, so
equations (\ref{eq:X2sSubstitution})-(\ref{eq:Y2cSubstitution}) reduce to
\begin{align}
\label{eq:X2sSubstitutionQS}
\tilde{X}_{2s}^{(0)} &= 
\frac{Y_{1c}}{2 Y_{1s}} \tilde{X}_{20}^{(0)}
-\frac{X_{1c}}{2 Y_{1s}} \tilde{Y}_{20}^{(0)},
\\
\tilde{X}_{2c}^{(0)} &= 
-\frac{1}{2} \tilde{X}_{20}^{(0)}
-\frac{s_G s_\psi X_{1c} \kappa}{4 Y_{1s}},
\\
\tilde{Y}_{2s}^{(0)} &= 
\frac{Y_{1c}^2 - Y_{1s}^2}{2 X_{1c} Y_{1s}} \tilde{X}_{20}^{(0)}
-\frac{Y_{1c}}{2 Y_{1s}} \tilde{Y}_{20}^{(0)}
-\frac{s_G s_\psi \kappa}{4 },
\\
\tilde{Y}_{2c}^{(0)} &= 
-\frac{Y_{1c}}{ X_{1c} } \tilde{X}_{20}^{(0)}
+\frac{1}{2} \tilde{Y}_{20}^{(0)}
-\frac{s_G s_\psi Y_{1c} \kappa}{4 Y_{1s}}.
\label{eq:Y2cSubstitutionQS}
\end{align}


\bibliographystyle{jpp}

\bibliography{figuresOfMeritNearAxis}

\end{document}